\documentclass[twocolumn,floatfix,prb,aps,showpacs,bibliography]{revtex4-1}
\usepackage{graphicx,amsmath,amssymb,color}
\usepackage{nicefrac}
\usepackage{url}
\usepackage[titletoc,title]{appendix}
\usepackage{color,amsmath,amssymb,amsfonts,graphicx,tabularx}
\usepackage{nicefrac}
\usepackage[titletoc,title]{appendix}
\usepackage[unicode=true,colorlinks=true]{hyperref}
\hypersetup{linkcolor=blue,citecolor=blue,urlcolor=blue}
\makeatletter
\newcommand*{\rom}[1]{\expandafter\@slowromancap\romannumeral #1@}
\makeatother

\newcommand{\ba}{\begin{eqnarray}}
\newcommand{\ea}{\end{eqnarray}}
\newcommand{\be}{\begin{equation}}
\newcommand{\ee}{\end{equation}}

\renewcommand{\vec}[1]{\mbox{\boldmath$#1$}}

\begin{document}

\title{Composite Fermions on a Torus}
\author{Songyang Pu$^{1}$, Ying-Hai Wu$^{2}$ and J. K. Jain$^{1}$}

\affiliation{$^{1}$Department of Physics, 104 Davey Lab, Pennsylvania State University, University Park, Pennsylvania 16802, USA \\ $^{2}$Max-Planck-Institut f{\"u}r Quantenoptik, Hans-Kopfermann-Stra{\ss}e 1, 85748 Garching, Germany}

\date{\today}

\begin{abstract} 
We achieve an explicit construction of the lowest Landau level (LLL) projected wave functions for composite fermions in the periodic (torus) geometry.  To this end, we first demonstrate how the vortex attachment of the composite fermion (CF) theory can be accomplished in the torus geometry to produce the ``unprojected" wave functions satisfying the correct (quasi-)periodic boundary conditions. We then consider two methods for projecting these wave functions into the LLL. The direct projection produces valid wave functions but can be implemented only for very small systems. The more powerful and more useful projection method of Jain and Kamilla fails in the torus geometry because it does not preserve the periodic boundary conditions and thus takes us out of the original Hilbert space. We have succeeded in constructing a modified projection method that is consistent with both the periodic boundary conditions and the general structure of the CF theory.  This method is valid for a large class of states of composite fermions, called ``proper states," which includes the incompressible ground states at electron filling factors $\nu=\frac{n}{2pn+ 1}$, their charged and neutral excitations, and also the quasidegenerate ground states at arbitrary filling factors of the form $\nu=\frac{\nu^*}{2p\nu^*+ 1}$, where $n$ and $p$ are integers and $\nu^*$ is the CF filling factor. Comparison with exact results known for small systems for the ground and excited states at filling factors $\nu=1/3$, 2/5 and 3/7 demonstrates our LLL-projected wave functions to be extremely accurate representations of the actual Coulomb eigenstates. Our construction enables the study of large systems of composite fermions on the torus, thereby opening the possibility of investigating numerous interesting questions and phenomena.
\end{abstract}

\maketitle

\tableofcontents
\section{Introduction}

The fractional quantum Hall effect (FQHE)\cite{Tsui82} is one of the most wonderful collective states discovered in nature, serving as a quintessential prototype for emergent topological order and triggering a wealth of novel physics and  concepts\cite{DasSarma07,Heinonen98,Jain07}. A central role in its explanation is played by explicit microscopic wave functions, which reveal the underlying physics, allow an explicit confirmation of this physics through comparisons to exact wave functions known for small systems, and enable calculation of observables that can be compared quantitatively with experimental measurements. 

In 1983 Laughlin constructed wave functions for the incompressible states at $\nu=1/(2p+1)$, $p$ integer, using the symmetric gauge of the planer geometry\cite{Laughlin83} in which the vector potential is given by $\vec{A}={1\over 2}B\vec{r}\times \hat{\vec{z}}$.  The Laughlin wave function was generalized by Haldane\cite{Haldane83} to the spherical geometry, in which electrons move on the surface of a sphere subject to a radial magnetic field.

Explicit wave functions for a broader class of fractional quantum Hall (FQH) states and their excitations were constructed within the composite fermion (CF) theory\cite{Jain89,Jain07}. Composite fermions are topological bound states of electrons and an even number ($2p$) of quantized vortices, often viewed as bound states of electrons and $2p$ magnetic flux quanta. They experience an effective magnetic field $B^*=B-2p\rho\phi_0$, where $B$ is the external magnetic field, $\rho$ is the electron or the CF density, and $\phi_0=hc/e$ is the flux quantum. Composite fermions form Landau-like levels in the effective magnetic field, called $\Lambda$ levels ($\Lambda$Ls), and have a filling factor $\nu^*$ given by $\nu=\nu^*/(2p\nu^*\pm 1)$. The CF theory provides a qualitative explanation of the phenomenology of the FQHE in the lowest Landau level (LLL). In particular, the FQHE of electrons at $\nu=n/(2pn\pm 1)$ is explained as the integer quantum Hall effect (IQHE) of composite fermions at CF filling factors $\nu^*=n$, with the $+$ ($-$) sign corresponding to the binding of positive (negative) vortices.

The Jain CF wave functions are constructed by ``composite-fermionizing" the known wave function of IQHE states of noninteracting electrons at filling factor $\nu^*$. The construction proceeds by first binding vortices to electrons to convert them into composite fermions, and then projecting the resulting wave function into the LLL.  For this purpose Laughlin's symmetric gauge of the planer geometry is the most convenient, because it allows a transparent definition for vortex attachment and a straightforward prescription for LLL projection \cite{Girvin84b,Jain89,Jain97,Jain97b}. However, the disk geometry is not very suitable for calculations of the bulk properties of the FQHE because of the presence of edges, which necessitates going to very large systems before the bulk behavior manifests itself, and also because each LL has an infinite degeneracy, thus making the definition of $n$ filled LL states ambiguous. Haldane's spherical geometry \cite{Haldane83} has proved more useful for practical calculations. Because of the compactness of this geometry, each Landau level (LL) has a finite degeneracy, and thus incompressible states are sharply defined.  The CF theory has been generalized to the spherical geometry \cite{Dev92,Wu93,Jain97,Jain97b}. Furthermore, a LLL projection method has been developed by Jain and Kamilla (JK) \cite{Jain97,Jain97b} that allows calculations for large numbers of composite fermions for both disk and spherical geometries. This has enabled the study of many states and phenomena that are not manifest in small systems, and also played an important role in carrying out detailed quantitative comparisons between the CF theory and experiment \cite{Jain07}.

Another important geometry for the study of the FQHE is the periodic, or the torus, geometry \cite{Yoshioka83}, which is the topic of this article. Already in 1985, Haldane and Rezayi \cite{Haldane85} generalized the Laughlin wave function to the torus geometry and showed that it has a $2p+1$-fold center-of-mass (CM) degeneracy. The periodic boundary conditions of the torus geometry are widely used in condensed matter physics, and in the context of FQHE, this geometry provides crucial information about the topological content of various FQH states through their ground state degeneracies. The torus geometry is also the most natural geometry for the study of crystal and stripe phases \cite{Rezayi00}, Hall viscosity \cite{Fremling14}, the thin torus limit \cite{Bergholtz05,Bergholtz06,Bergholtz08b}, mapping into spin models \cite{Nakamura11,Wang12f}, entanglement properties \cite{Sterdyniak11}, and edge structure \cite{Lopez99}. This geometry is necessary for studying FQHE for interacting particles on a Hofstadter lattice \cite{Hofstadter76,Thouless82}. In recent years, there has been interest in the feasibility of ``fractional Chern insulators," which refer to FQH states of interacting particles in Haldane-type lattice models \cite{Haldane88} where the net magnetic field through a unit cell is zero. Various articles have investigated fractional Chern insulators by numerical diagonalization in the torus geometry \cite{ShengDN2011,Regnault11,Wang11,Bernevig12,Wu12b,Scaffidi2012,LiuZ2013}, and wave functions for certain FQH states have been constructed \cite{Qi11,Wu12e,WuYL2013}. Finally, the torus geometry is very useful for studying competition between different candidate states at a given filling factor. In the spherical geometry, such candidate states often occur at different $N_\phi$ for a finite system (although $\nu=\lim_{N\rightarrow\infty} N/N_\phi$ is the same for all of them), making it difficult to carry out a direct comparison or to study the phase transition between them. (Here $N$ is the number of electrons and $N_\phi$ is the number of flux quanta $\phi_0=hc/e$ passing through the sample.)

\begin{figure}
\includegraphics[width=0.6\textwidth]{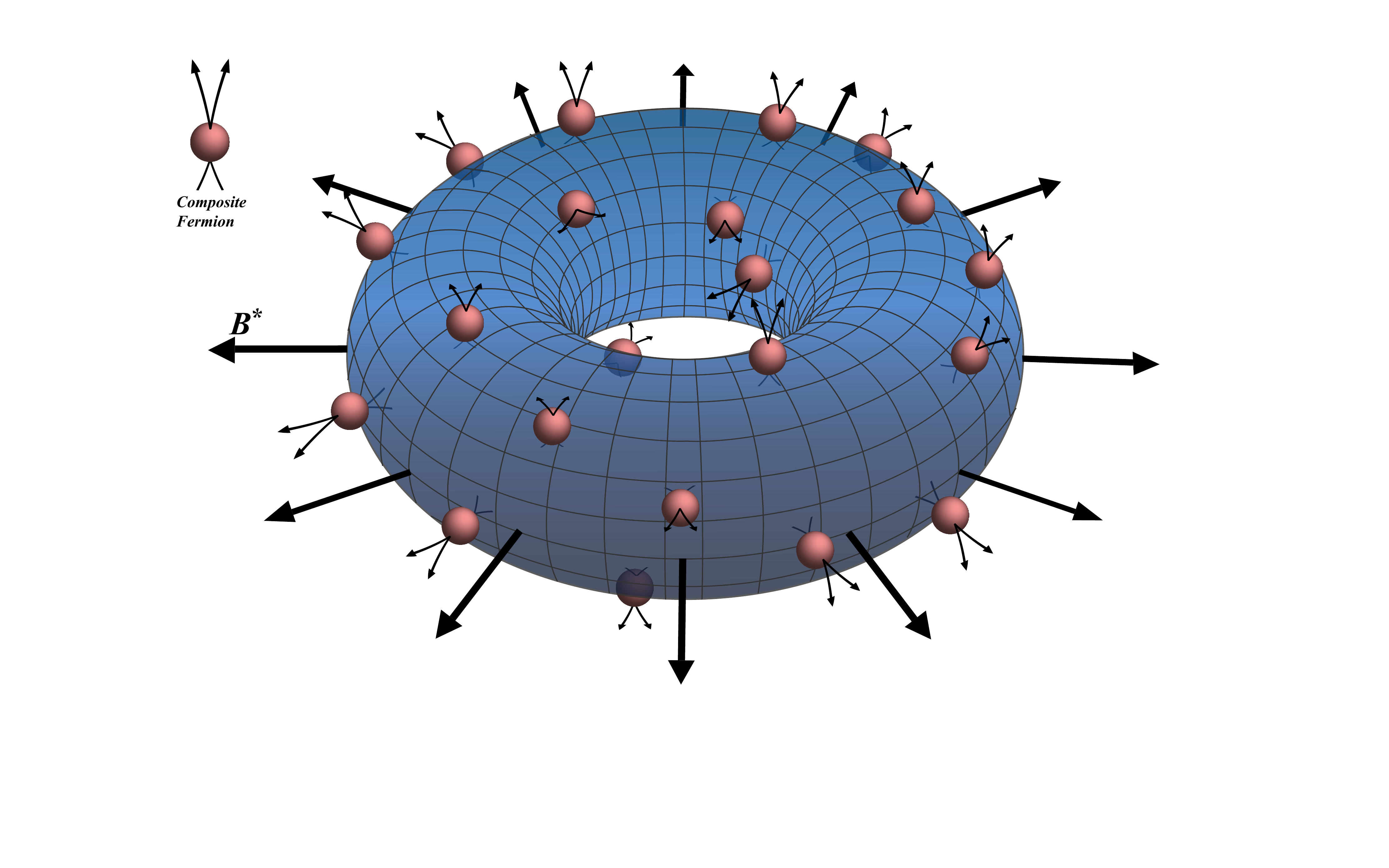}
\caption{Composite fermions on a torus.}
\label{TorusShape}
\end{figure}

It would therefore be extremely useful to have explicit wave functions of composite fermions on a torus (Fig.~\ref{TorusShape}) for the investigation of general FQH states and their excitations. A generalization of the CF theory to the torus geometry has proved nontrivial, however. One of the stumbling blocks appears to be that the natural gauge for the torus geometry is the Landau gauge, whereas the most natural gauge for composite fermionization is the symmetric gauge. In the symmetric gauge, vortex attachment is accomplished by multiplication by the factor $\prod_{j<k}(z_j-z_k)$ and LLL projection amounts, essentially, to the replacement $\bar{z}\rightarrow 2\partial/\partial z$, where $z=x+iy$ denotes the electron coordinates; analogous simple forms are not available in the Landau gauge. Additionally, it is not immediately clear how to represent derivatives in the periodic geometry, which are an integral part of the LLL-projected Jain wave functions.

One may wonder why it should be easier to construct, in the torus geometry, the Laughlin wave function than to construct the wave functions for other FQH states. The reason is that the Laughlin wave function has a Jastrow form with a simple analytic structure: In this wave function, {\em all} zeros of a given particle coordinate sit exactly on other particles; i.e., there are no wasted zeros.  The Laughlin wave function is thus fully determined, with the LLL restriction, by specifying the short distance behavior as two particles are brought together. Ensuring the short distance behavior along with the correct periodic boundary conditions is sufficient to uniquely identify the Jastrow form for the Laughlin wave function on the torus, modulo the CM degree of freedom \cite{Haldane85}. It is evident that this principle cannot be extended for the construction of wave functions for the general FQH states, because they do not have a Jastrow form and are not uniquely determined by their short distance behavior. For example, with the exception of the ground state at $\nu=1/3$, all LLL wave functions at filling factors of the form $\nu=n/(2n+1)$ {\em must} vanish as a single power of the distance between two particles as they are brought close. Thus there is only one zero on each particle, with the remaining zeros distributed in a very complex fashion \cite{Graham03}. Even the wave function of a single quasiparticle of the 1/3 state was left as an open problem in Ref.~\onlinecite{Haldane85}. The CF theory circumvents this issue by approaching the problem from a different paradigm, which shows that the seemingly complex LLL wave functions are ``adiabatically connected to," and LLL projections of, simpler wave functions that reveal the physics of the more general states in terms of composite fermions occupying $\Lambda$Ls.

Significant progress has been made in writing wave functions for general FQH states in the torus geometry based on a conformal field theory (CFT) formulation of composite fermions \cite{Cristofano95,Cristofano04,Hermanns08,Bergholtz08,Hansson09,Hansson09b,Marotta09,Marotta10,Rodriguez12b,Cappelli13,Hermanns13,Fremling14,Estienne15,Chen17} and an explicit construction of their wave functions as CFT correlators  \cite{Hansson07,Hansson07a}. Hermanns {\em et al.} \cite{Hermanns08} constructed wave functions for the ground states at $\nu=n/(2pn+1)$ in the torus geometry. They demonstrated that the resulting wave function for the 2/5 state has a high overlap with the exact Coulomb wave function. This construction was generalized to arbitrary fractions by Bergholtz {\em et al.}\cite{Bergholtz08}. Hansson {\em et al.} \cite{Hansson09b,Hansson09} constructed wave functions for the quasiparticles of both Abelian and non-Abelian FQH states, again with guidance from CFT.  More recently, Hermanns\cite{Hermanns13} constructed the ground state wave functions for  $\nu=n/(2pn+1)$ following the standard approach\cite{Jain89,Jain07}, and demonstrated, for small systems, that projection into the LLL produces wave functions that have very high overlap with the exact Coulomb eigenstates. Quasiparticles of the Laughlin state were also considered by Greiter {\em et al.}\cite{Greiter16}. Fremling {\em et al.}\cite{Fremling16} have developed an energy projection method to produce LLL CF wave functions in the torus geometry. 
These advances notwithstanding, exact diagonalization has remained the primary method for studying the general FQH states in the torus geometry, because the currently available wave functions are not easy to work with and cannot be evaluated for systems larger than those accessible to exact diagonalization studies. 

We present in this work a different construction for the LLL wave functions of composite fermions. The advantage of our method is that we can construct wave functions for a large class of ground and excited states at arbitrary filling factors of the form $\nu=\nu^*/(2p\nu^*+1)$, and also evaluate them on the computer for much larger systems than possible in exact diagonalization studies. We give here a brief outline of our method, which should be useful for the reader who is not interested in the technical details. More complete derivations and explicit expressions can be found in the subsequent sections and appendices.

We consider a torus defined by two edges of the parallelogram $\xi_1=L_1$ and $\xi_2=L_1\tau$, where $\tau$ is a complex number that specifies the geometry of the torus (see Fig. \ref{TorusShape}). The magnetic field must be chosen so that an integer number $N_\phi$ flux quanta pass through the system. A crucial step below is to express the single particle wave functions in the torus geometry in the symmetric gauge\cite{Greiter16}, reviewed in Sec. \ref{Single particle functions}. The single-particle wave functions are chosen to satisfy the boundary conditions
\begin{equation}
\label{pbc}
\begin{gathered}
t(L_1)\psi(z)=e^{i\phi_1}\psi(z)\\
t(L_1\tau)\psi(z)=e^{i\phi_\tau}\psi(z)
\end{gathered}
\end{equation}
where $t(L_1)$ and $t(L_1\tau)$ are magnetic translation operators. The phases $\phi_1$ and $\phi_\tau$ define the  Hilbert space. 
It is convenient to write the single particle wave functions as
\be
\psi(z)=e^{\frac{z^2-|z|^2}{4l^2}}f(z)
\ee
where $l=\sqrt{\hbar c/eB}$ is the magnetic length.
The wave function of $n$ filled LLs is denoted as
\be
\Psi_n\equiv e^{\sum_i\frac{z_i^2-|{z}_i|^2}{4l^2}}\chi_n(f_i(z_j))
\ee
where $\chi_n(f_i(z_j))$ is a Slater determinant formed from the single particle wave functions $f_i(z_j)$, where the subscript $i$ denotes collectively the quantum numbers (LL index, momentum) of the single-particle state. 
In particular, the wave function of one filled LL, $\Psi_1$, assumes the simple form 
\be
\Psi_1[z_i,\bar{z}_i]\sim e^{\sum_i\frac{z_i^2-|z_i|^2}{4l^2}}F_1(Z)\prod_{j<k}\theta \left(\frac{z_j-z_k}{L_1}|\tau \right)
\label{LLLstate}
\ee
where $\theta$ is the odd Jacobi theta function \cite{Mumford07} and 
\be
\label{CMZ}
Z=\sum_{i=1}^N z_i
\ee
is the CM coordinate for a system of $N$ electrons. 

In Sec. \ref{products of single particle wave functions} we show how a product of three single particle wave functions produces, with an appropriate choice of boundary conditions for each factor, a valid wave function in our Hilbert space. The magnetic field of the product is the sum of the magnetic fields of the individual factors. It thus follows that the standard unprojected Jain wave functions
\be
\Psi^{\rm unproj}_{n\over 2pn+1}=\Psi_n \Psi_1^{2p}
\ee
are legitimate wave functions, where $\Psi_n$ is wave function of $n$ filled LLs in an effective magnetic field corresponding to magnetic flux
\be
\label{number of flux}
N^*_\phi=N_\phi-2p N
\ee
where $N_\phi$ is the physical magnetic flux,
and $\Psi_1$ is constructed at magnetic flux $N^{(\nu=1)}_\phi=N$.
 The states $\Psi^{\rm unproj}$ are in general not confined to the LLL, however, and ought to be projected into the LLL to calculate quantities appropriate for the large magnetic field limit where admixture with higher LLs is negligible.

The use of symmetric gauge allows us to accomplish the LLL projection exactly as in the disk geometry, i.e., by moving all $\bar{z}$'s to the left and replacing $\bar{z}\rightarrow 2l^2\partial /\partial z$ with the understanding that the derivatives do not act on the Gaussian factor $e^{-|z|^2/4l^2}$.  This produces the LLL projected wave function 
\begin{widetext}
\be
\Psi_{n\over 2pn+1} = e^{\sum_i\frac{z_i^2-|{z}_i|^2}{4l^2}} \chi_n[\hat{f}_i(\partial/\partial z_j,z_j)]F_1^{2p}(Z) \prod_{j<k}\left[\theta \left(\frac{z_j-z_k}{L_1}|\tau \right)\right]^{2p}
\label{JK1}
\ee
\end{widetext}
where the operator $\hat{f}_{i}(\partial/\partial z_j,z_j)$ is obtained from the single particle wave function $f_{i}(\bar{z}_j,z_j)$ by moving $\bar{z}_j$ to the left and making the replacement $\bar{z}_j\rightarrow 2 l^2 \partial/\partial z_j+z_j$. 
This is analogous to the original method for projection, called ``direct projection" in the disk and spherical geometries \cite{Dev92,Dev92a,Wu93}, and the resulting wave functions are equivalent, modulo gauge choice, to those obtained by Hermanns\cite{Hermanns13}. The direct projection corresponds to expanding the unprojected wave function in terms of the Slater determinant basis functions and keeping only the part residing in the LLL. This projection originally played a crucial role in establishing the validity of the CF theory, but is not useful for practical calculations, because it allows projection for only small systems\cite{Dev92,Dev92a,Wu93,Wu95}. The reason is that one needs to keep track of all individual LLL Slater determinant basis functions, the number of which grows exponentially with the system size and soon becomes too large to store. A more useful form for LLL projection was obtained by JK\cite{Jain97,Jain97b}, which can be implemented for large systems of composite fermions, allowing a determination of thermodynamic limits for many quantities of interest. Both the direct and the JK projection methods produce very accurate, though not identical, LLL wave functions.

We implement the JK projection in the following fashion. We show in Appendix \ref{proof that operator commute with F(Z)} that in Eq.~\ref{JK1} the CM part $F_1^{2p}(Z)$ can be commuted through $\chi_n$. Then, in the spirit of the JK projection method (briefly reviewed in Sec.~\ref{JKreview}), we write 
\be
\Psi^{\rm JK}_{n\over 2pn+1}= e^{\sum_i\frac{z_i^2-|z_i|^2}{4l^2}}F_1^{2p}(Z) \chi_n[\hat{f}_{i}(\partial/\partial z_j,z_j) J^p_j]
\ee
where 
\be 
J_j=\prod_{k (k\neq j)}\theta \left(\frac{z_j-z_k}{L_1}|\tau \right)
\ee
In Sec. \ref{Failure of the JK projection method} we show that the JK projection method fails in the torus geometry, because it does not preserve the periodic boundary conditions and thus takes us out of our original Hilbert space. 

The principal achievement of our work is to show that, for the so-called ``proper states" defined below,  it is possible to construct a modified projection method that produces LLL wave functions that have the CF structure and also satisfy the correct boundary conditions. In essence, we derive a closely related operator $\hat{g}$ such that 
\be
\Psi_{n\over 2pn+1} = e^{\sum_i\frac{z_i^2-|{z}_i|^2}{4l^2}} F_1^{2p}(Z)\chi_n[\hat{g}_{i}(\partial/\partial z_j,z_j) J^p_j]
\label{ghat}
\ee
satisfies the correct boundary conditions. It is shown in Sec. \ref{Modified LLL projection method} and Appendix \ref{proof modified JK} that $\hat{g}_{i}(\partial/\partial z_j,z_j)$ is obtained from $\hat{f}_{i}(\partial/\partial z_j,z_j)$ by making the replacement $\partial/\partial z_k\rightarrow 2\partial/\partial z_k$ for all derivatives acting on $J^p_j$. We note that $F_1^{2p}(Z)$ does not account for the entire CM coordinate dependence of the wave function.

It is known from general considerations that there are $2pn+1$ degenerate eigenstates at $\nu=n/(2pn+1)$, which are related by CM translation operator. The natural wave function obtained in the CF theory is in general not an eigenstate of the CM translation operator, but is a specific linear superposition of the $2pn+1$ degenerate ground states. Appendix \ref{CM deg} shows how, within our approach, we can construct $2pn+1$ eigenstates of the CM translation operator. 

\begin{figure}[!h]
\includegraphics[width=0.22\textwidth, height=0.13\textwidth]{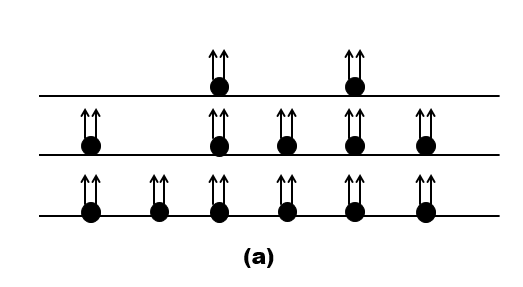} 
\includegraphics[width=0.22\textwidth, height=0.13\textwidth]{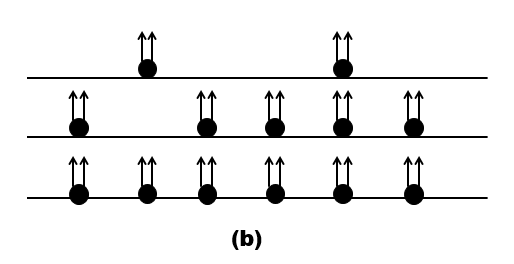}
\caption{(a) An example of a ``proper state."  No composite fermion in any $\Lambda$ level has a hole directly underneath it in any $\Lambda$ level. (A composite fermion is depicted as an electron bound to two flux quanta represented by vertical arrows.) (b)  An example of a state that is not proper. 
}
\label{proper}
\end{figure}

\begin{figure}[!h]
\includegraphics[width=7.5CM,height=5.0CM]{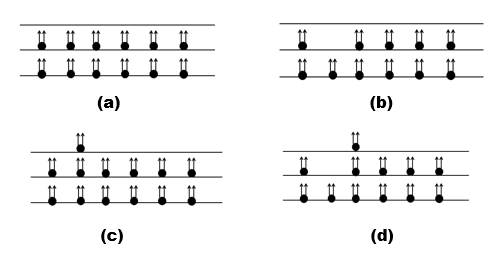} 
\caption{Examples of certain important types of proper states: (a) the 2/5 ground state, (b)  a CF quasihole at $\nu=2/5$, (c) a CF quasiparticle at $\nu=2/5$, and (d) a neutral CF exciton at $\nu=2/5$.
}
\label{four states}
\end{figure}

We further show that our method provides legitimate wave functions for a much broader class of states, which we term ``proper states." A proper state is defined by the condition that if the orbital of a given ``momentum" quantum number is occupied in the $n$th $\Lambda$L, then it is occupied in all of the lower $\Lambda$Ls. An example of a proper state is shown in Fig.~\ref{proper}, along with a state that is not proper. Proper states include (i) the ground states at $\nu=n/(2pn+1)$;  (ii) CF quasiholes, which contain $n$ $\Lambda$Ls fully occupied except for a single hole in the $n$th $\Lambda$L; (iii) CF quasiparticles, which contain a single composite fermion in the $(n+1)^{\rm st}$ $\Lambda$L with the lowest $n$ $\Lambda$Ls fully occupied; (iv) neutral excitations, which contain a CF-particle hole pair, provided that the particle is not directly above the hole. These are depicted in Fig.~\ref{four states}. In all of these cases, we first construct the Slater determinant $\chi_{\nu^*}$ for the corresponding state at $\nu^*$,  and composite-fermionize it to obtain 
\be
\Psi_{\nu^*\over 2p\nu^*+1} = e^{\sum_i\frac{z_i^2-\bar{z}_i^2}{4l^2}}F_1^{2p}(Z) \chi_{\nu^*}(\hat{g}_{i}(\partial/\partial z_j,z_j) J^p_j)
\label{modWF}
\ee
We show in Appendix~\ref{Negative} that the construction is also valid for the Jain states at $\nu=\nu^*/(2p\nu^*-1)$ requiring negative vortex (or flux) attachment; however, we will not consider these states explicitly because their evaluation is much more complicated than that of the states $\nu=\nu^*/(2p\nu^*+1)$.

The evaluation of the wave functions in Eq.~\ref{modWF} does not require expansion into Slater determinant basis functions, and thus can be performed for very large systems. For illustration, we show below results for up to $N=40$ particles. We have not made any attempt to ascertain the largest $N$ for which calculations are possible.

The fact that we have constructed LLL wave functions does not guarantee, by any means, that these wave functions are accurate representations of the actual eigenstates of electrons interacting via the repulsive Coulomb interaction. That must be checked by explicit calculation. We demonstrate the quantitative accuracy of our LLL-projected wave functions by comparison with exact results known for small systems. In particular, in Sec. \ref{Testing the validity of the LLL projected sates} we calculate the Coulomb energies of our wave functions for the CF quasiparticle at 1/3 and for the ground states, CF quasiparticles, and CF quasiholes at $\nu=2/5$ and 3/7. These energies are very close to the Coulomb energies obtained from exact diagonalization, establishing the quantitative validity of our torus wave functions. 

A remark on units is in order. We will quote the energies in units of $e^2/\epsilon l$, where $l$ is the magnetic length and $\epsilon$ is the dielectric function of the background material. We will {\em not} use, as is the general practice, the magnetic length $l$ as the unit of length, but will explicitly display it.

\section{Single particle wave functions on torus} \label{Single particle functions}

A torus is topologically equivalent to a parallelogram with periodic boundary conditions. We define the two edges of the parallelogram to be $\xi_1=L_1,\xi_2=L_1\tau$, where $\tau$ is a complex number that represents the aspect ratio of the torus. The magnetic field is perpendicular to the plane of parallelogram $\vec{B}=-B\hat{\vec{z}}$. We choose the symmetric gauge $\vec{A}=\frac{1}{2}B\vec{r}\times\hat{\vec{z}}$, which would be crucial for accomplishing LLL projection. In this subsection we describe the single particle wave functions following the conventions in Ref.~\onlinecite{Greiter16}. 

We use $z=x+iy,\bar{z}=x-iy$ as the coordinates for particles. To describe the cyclotron and guiding-center variables, we define two sets of ladder operators:
\begin{equation}
\label{ladder}
\begin{gathered}
a=\sqrt{2}l(\partial_{\bar{z}}+\frac{1}{4l^2}z); a^\dagger=\sqrt{2}l(-\partial_{z}+\frac{1}{4l^2}\bar{z})\\
b=\sqrt{2}l(\partial_{z}+\frac{1}{4l^2}\bar{z}); b^\dagger=\sqrt{2}l(-\partial_{\bar{z}}+\frac{1}{4l^2}z)
\end{gathered}
\end{equation} 
These satisfy 
\be
[a,a^\dagger]=[b,b^\dagger]=1
\ee
and all other commutators vanish.
In terms of the ladder operators, the single-particle Hamiltonian can be recast as
\be
H=\frac{1}{2M}\left(\vec{p}+\frac{e}{c}\vec{A}\right)^2=\hbar \omega_{c}\left(a^\dagger a+\frac{1}{2}\right)
\ee
where $\omega_{c}=eB/Mc$ is the cyclotron frequency and $M$ is the electron mass. 

On torus geometry, the wave functions are taken to satisfy the (quasi)periodic boundary conditions: 
\begin{equation}
\label{pb}
\begin{gathered}
t(L_1)\psi(z)=e^{i\phi_1}\psi(z)\\
t(L_1\tau)\psi(z)=e^{i\phi_\tau}\psi(z)
\end{gathered}
\end{equation}
where the magnetic translation operator is defined as 
\be
\label{magnetic translation operator}
t(\xi)=e^{-\frac{i}{2l^2}\hat{\vec{z}}\cdot(\vec{\xi}\times\vec{r})}T({\xi})=e^{\frac{1}{\sqrt{2}l}(\xi b-\bar{\xi} b^{\dagger})}
\ee
In our convention, $t$ represents the magnetic translation operator and $T$ represents the usual translation operator
\be
T(\xi)=exp\left(\xi \partial_z+\bar{\xi} \partial_{\bar{z}}\right).
\ee 
The following relationship between $t$ and $T$ will be very useful later: 
\begin{equation}
\label{tilde t}
{t}(\alpha L_1)e^\frac{z^2-|z|^2}{4l^2}=e^\frac{z^2-|z|^2}{4l^2} T(\alpha L_1)
\ee
\be
{t}(\alpha L_1\tau) e^\frac{z^2-|z|^2}{4l^2}=e^\frac{z^2-|z|^2}{4l^2}e^{i\alpha \pi N_\phi(2z/L_1+\alpha \tau)}T(\alpha L_1\tau)
\ee
where $\alpha$ is a real number between $0$ and $1$.
It is evident from Eq.~\ref{magnetic translation operator} that the magnetic translation operators commute with the ladder operators $a$ and $a^\dagger$:
\be
\left[t,a\right]=\left[t,a^\dagger \right]=0
\ee

The commutation relation $[t(L_1),t(L_1\tau)]=0$ imposes the condition that the number of flux quanta through the surface of the torus, i.e., 
\begin{equation}
\label{quantization of tau}
 N_\phi=\frac{L_1^2 {\rm Im}(\tau)B}{\phi_0}
\end{equation}
is an integer. Here a flux quantum is defined as $\phi_0=hc/e$.

\subsection{Lowest Landau level}

We first review the construction of single-particle wave functions in the LLL in the symmetric gauge, closely following Greiter {\em et al.} \cite{Greiter16}.  For this purpose it is convenient to write 
\begin{equation}
\label{single wf with theta form}
\psi_1(z)=e^{\frac{z^2-|z|^2}{4l^2}}f_1(z)
\ee
where the subscript $n=1$ refers to the LLL. (We stress that our convention is different from most other literature, where the LLL is defined as $n=0$.) Combining Eq.~\ref{pb} with Eq.~\ref{single wf with theta form}, and making use of Eq.~\ref{tilde t}, we get the periodic boundary conditions for $f_1(z)$:
\begin{equation}
\begin{gathered}
\label{pb f(z) 1}
\frac{T(L_1)f_1(z)}{f_1(z)}=\frac{f_1(z+L_1)}{f_1(z)}=e^{i\phi_1}\\
\frac{T(L_1\tau)f_1(z)}{f_1(z)}=\frac{f_1(z+L_1\tau)}{f_1(z)}=e^{i(\phi_\tau-\pi N_\phi(2z/L_1+\tau))}
\end{gathered}
\end{equation}
The solutions for Eq.~\ref{pb f(z) 1} are given by \cite{Haldane85}:  
\be
\label{f(z)}
f_1(z)=e^{ikz} \prod_{\nu =1}^{N_\phi} \theta(z/L_1-w_{\nu}|\tau)
\ee
where $\theta(z|\tau)$ is the odd Jacobi theta function\cite{Mumford07}:
\be
\theta(z|\tau)=\sum_{n=-\infty}^{\infty}e^{i\pi \left(n+\frac{1}{2}\right)^2\tau}e^{i2\pi \left(n+\frac{1}{2}\right)\left(z+\frac{1}{2}\right)}
\end{equation}
The odd Jacobi theta function is variously denoted as $\theta_{\frac{1}{2},\frac{1}{2}}(z|\tau)$ or $\theta_{1}(z|\tau)$ in the literature. For simplicity we shall suppress the subscript and use $\theta(z|\tau)$, because we do not use other types of Jacobi theta functions in this work.

The dimension of the Hilbert space in the LLL is $N_\phi$. To form a complete and orthogonal basis for this Hilbert space, we make the following choice for $\psi_1^{(n)}(z)$:
\begin{equation}
\label{lowest level single wf}
\begin{gathered}
\psi_1^{(n)}(z)=e^{\frac{z^2-|z|^2}{4l^2}}f_1^{(n)}(z)\\
f_1^{(n)}(z)=e^{ik^{(n)}z} \prod_{\nu =1}^{N_\phi} \theta(z/L_1-w^{(n)}_{\nu}|\tau),\ n=0,1,\dots , N_\phi-1\\
k^{(n)}=\frac{\phi_1-\pi N_\phi+2\pi n}{L_1},\\
w_\nu^{(n)}=\frac{1}{2\pi N_\phi}\left(\phi_\tau-\phi_1 \tau-\pi N_\phi(2-\tau)-2\pi n\tau+\pi+2\pi (\nu -1)\right)
\end{gathered}
\end{equation} 
The relations 
\begin{widetext}
\be
\begin{gathered}
\label{index n}
t\left(\frac{L_1}{N_\phi}\right)\psi_1^{(n)}(z)=-e^{i\frac{k^{(n)}L_1}{N_\phi}}\psi_1^{(n)}(z)\\
t\left(\frac{L_1\tau}{N_\phi}\right)\psi_1^{(n)}(z)=e^{i\tau \left(\pi+\frac{k^{(n)}L_1}{N_\phi}\right)}\psi_1^{(n+1)}(z),n=0\ldots N_\phi-2\\
t\left(\frac{L_1\tau}{N_\phi}\right)\psi_1^{(N_\phi-1)}(z)=e^{i \left(\pi \tau +\frac{k^{(N_\phi-1)}L_1 \tau}{N_\phi} +\phi_\tau-\phi_1 \tau \right)}\psi_1^{(0)}(z),n=N_\phi-1\\
\end{gathered}
\ee
\end{widetext}
show that $\psi_1^{(n)}(z)$ are eigenfunctions of $t\left(\frac{L_1}{N_\phi}\right)$, and  are related to one another by application of $t\left(\frac{L_1\tau}{N_\phi}\right)$. 
We will call $k^{(n)}$ the ``momentum" of $\psi_1^{(n)}(z)$. 

\subsection{Higher Landau levels}

In the CF construction of the FQHE states, we need wave functions for higher LLs.  
Using 
\be
\label{higher f}
a^\dagger e^{\frac{z^2-|z|^2}{4l^2}}=e^{\frac{z^2-|z|^2}{4l^2}}\sqrt{2}l\left(\frac{\bar{z}-z}{2l^2}-\partial_z\right),
\ee
the single particle wave function in the $m$th LL is given by 
\be 
\psi_m^{(n)}(z,\bar{z})=e^{\frac{z^2-|z|^2}{4l^2}}{(a_f^\dagger)^{m-1}\over \sqrt{(m-1)!}}f_1^{(n)}(z)
\ee
where 
\be
\label{af}
a_f^\dagger\equiv \sqrt{2}l\left(\frac{\bar{z}-z}{2l^2}-\partial_z\right)
\ee
For future reference, the single particle wave functions in the second and third LLs are:
\be
\label{single wf LL2}
\psi_2^{(n)}(z,\bar{z})=e^{\frac{z^2-|z|^2}{4l^2}}\sqrt{2}l\left(\frac{\bar{z}-z}{2l^2}f_1^{(n)}(z)-\frac{\partial f_1^{(n)}(z)}{\partial z}\right)
\ee
\begin{multline}
\label{single wf LL3}
\psi_3^{(n)}(z,\bar{z})=e^{\frac{z^2-|z|^2}{4l^2}}\sqrt{2} l^2\left[\left(\frac{\bar{z}-z}{2l^2}\right)^2f_1^{(n)}(z)\right. \\ \left. -\frac{\bar{z}-z}{l^2}\frac{\partial f_1^{(n)}(z)}{\partial z}+\frac{1}{2l^2}f_1^{(n)}(z)+\frac{\partial^2 f_1^{(n)}(z)}{\partial z^2}\right]
\end{multline}
That $\psi_m^{(n)}(z,\bar{z})$ satisfies periodic boundary conditions of Eq.~\ref{pb} follows because $a^\dagger$ commutes with the magnetic translation operators.
For the same reason, $\psi_m^{(n)}(z,\bar{z})$ also satisfies Eq.~\ref{index n}, and thus is labeled by the momentum $k^{(n)}$. It should be clear that the dimension of the Hilbert space is $N_\phi$ in all LLs. (This should be contrasted with the spherical geometry, for which the dimension increases by two for each successive LL.)

In what follows below, we will omit the overall normalization factors for various wave functions. These are not important when we consider many-body wave functions that are derived from a single Slater determinant, as will be the case in this article. 

With apologies, we note that the symbols $n$ and $m$ will be used to label both the momentum and the LL or $\Lambda$L indecies. In the wave function $\psi_m^{(n)}$ or $f_m^{(n)}$, the lower index refers to the LL or the $\Lambda$L index and the upper to the momentum. We hope this will not lead to any confusion.

\subsection{Wave function for one filled LL}

With the knowledge of the single-particle wave functions we can construct many particle wave functions as linear superpositions of Slater determinants. In particular, the ground state wave function $\Psi_n[z_i,\bar{z}_i]$ at filling $\nu=n$ is a single Slater determinant. Of special relevance below will be the wave function $\Psi_1$ of one filled LL:
\be
\label{LLL filled}
\Psi_1[z_i,\bar{z}_i]=e^{\sum_i\frac{z_i^2-|z|_i^2}{4l^2}}\chi_1[f_i(z_j)]
\ee
\be
\label{LLL filled2}
\chi_1[f_i(z_j)]=
\begin{vmatrix}
f_1^{(0)}(z_1)&\ldots&f_1^{(0)}(z_N) \\
f_1^{(1)}(z_1)&\ldots &f_1^{(1)}(z_N)\\
\vdots&\vdots&\vdots \\
f_1^{(N_\phi-1)}(z_1)&\ldots&f_1^{(N_\phi-1)}(z_N) \\
\end{vmatrix}
\ee
As shown in Appendix \ref{Laughlin wave functions}, $\Psi_1$ has the simple form
\be
\Psi_1[z_i,\bar{z}_i]={\cal N}e^{\sum_i\frac{z_i^2-|z|_i^2}{4l^2}}F_1(Z)\prod_{j<k}\theta \left(\frac{z_j-z_k}{L_1}|\tau \right)
\label{LLL filled3}
\ee
where ${\cal N}$ is a normalization factor.
In particular, $\chi_1$ is a product of a factor that depends only on the CM coordinate defined in Eq.~\ref{CMZ} 
and a factor that contains only the relative coordinates. The last expression in the above equation follows because it is the only function that depends only on $z_i$'s, vanishes as a single power of the distance when two particles are brought together, and is consistent with the periodic boundary conditions.  Appendix \ref{Laughlin wave functions} shows that the wave functions in Eqs.~\ref{LLL filled2} and \ref{LLL filled3} have the same behavior under CM translation. 

\section{Composite fermions}

In this section, we construct wave functions for low energy states at arbitrary filling factors of the form $\nu=\frac{\nu^*}{2p\nu^*+1}$ in terms of composite fermions at filling $\nu^*$. Our construction is valid for all ``proper" states defined in the introduction, which include the incompressible ground states at $\nu=n/(2pn+1)$, their charged and neutral excitations (except the neutral exciton in which the excited CF particle is directly above the CF hole left behind), and the quasidegenerate ground states at arbitrary fillings. For this purpose, we first prove that the product of single particle wave functions preserves the periodic boundary conditions. Then we construct the unprojected Jain wave functions and their Direct projection into the LLL. We finally show that the standard JK projection method fails for the torus geometry, but a modified projection method yields legitimate LLL wave functions for all proper states. In the following section, we explicitly evaluate the Coulomb energies of ground and excited states at $\nu=\frac{1}{3}$, $\nu=\frac{2}{5}$, and $\nu=\frac{3}{7}$ and find that they are extremely close to the corresponding exact energies.

\subsection{Products of single particle wave functions}
\label{products of single particle wave functions}

The general wave functions for composite fermions are the products of Slater determinants. Therefore, we begin by asking what periodic boundary conditions should be imposed on each factor to ensure the product satisfies the right periodic boundary conditions. To this end, we consider products of single particle wave functions:
\ba
\label{single product}
\psi(z,\bar{z})&=&\prod_i\psi^{(i)}(z,\bar{z}) \nonumber \\
&=&\prod_{i}e^{\frac{z^2-|z|^2}{4l^{(i)2}}}f^{(i)}(z,\bar{z})\nonumber \\
&=& e^{\frac{z^2-|z|^2}{4l^{2}}} \prod_{i} f^{(i)}(z,\bar{z})
\ea
The magnetic length $l$ of the product is related to the magnetic lengths of the individual factors as
\be
\frac{1}{l^2}=\sum_{i}\frac{1}{l^{(i)2}}
\ee
which implies that 
\be
\label{sum of magnetic field}
\begin{gathered}
N_\phi=\sum_i N_\phi^{(i)}\\
\end{gathered}
\ee

The boundary conditions for $\psi(z,\bar{z})$ with phases $\phi_1$ and $\phi_\tau$ translates into
\begin{widetext}
\be
\label{pb f(z)}
\begin{gathered}
\frac{T(L_1)[\prod f^{(i)}(z)]}{[\prod f^{(i)}(z)]}=\prod_{i}e^{i\phi_1^{(i)}}=e^{i\phi_1}\\
\frac{T(L_1\tau)[\prod f^{(i)}(z)]}{[\prod f^{(i)}(z)]}=\prod_{i}e^{i[\phi_\tau^{(i)}-\pi N_\phi^{(i)}(2z/L_1+\tau)]}=e^{i[\phi_\tau-\pi N_\phi(2z/L_1+\tau)]}\\
\end{gathered}
\ee
\end{widetext}
where $\phi_1^{(i)}$ and $\phi_\tau^{(i)}$ are the phases for the boundary conditions on the individual factors $f^{(i)}(z)$. 
Equations~\ref{pb f(z)} are satisfied provided we set
\be
\label{pbs for single product}
\begin{gathered}
\phi_1=\sum_i\phi_1^{(i)}\\
\phi_\tau=\sum_i\phi_\tau^{(i)}
\end{gathered}
\ee
and also make use of Eq.~\ref{sum of magnetic field}.

The above proof works for a product of any number of single particle wave functions. As shown in Appendix~\ref{Negative}, the product also satisfies the correct boundary conditions if the first single particle wave function is evaluated at a ``negative'' magnetic field, i.e., $\psi^{(1)}(z)$ is replaced by its complex conjugate. (We thank Mikael Fremling for pointing out that a similar construction works in the Landau or $\tau$ gauge, which helped us eliminate an error in an earlier version of the manuscript.)
We will consider only the states at $\nu=\nu^*/(2\nu^*+1)$ in what follows because the LLL projection for states at $\nu=\nu^*/(2\nu^*-1)$ is much harder to evaluate.

\subsection{Unprojected wave functions}

A composite fermion is the bound state of an electron and even number $2p$ of quantized vortices. For the ground states of $\nu=\frac{n}{2pn+1}$, we write the unprojected wave functions:
\be
\label{CF product}
\Psi^{\rm unproj}_{n\over 2pn+1}=\Psi_n \Psi_1^{2p}
\ee 
where $\Psi_n$ is the wave function of $n$ filled LLs at the effective flux quanta $N^*_\phi=N/n$ and $\Psi_1$ is the wave function of $1$ filled LL at the effective flux quanta $N^{(1)}_\phi=N$. The product wave function $\Psi^{\rm unproj}_{n\over 2pn+1}$ occurs at flux 
\be
\label{number of flux}
N_\phi=N^*_\phi+2p N
\ee
and thus corresponds to 
\be
\nu={N\over N_\phi}={n\over 2pn+1}
\label{fillingfactor}
\ee
Recalling that the translation operators for different particles commute, the results of the previous section regarding products of single particle wave functions imply that $\Psi^{\rm unproj}_{n\over 2pn+1}$ satisfies the correct boundary conditions.
Because the number of states in each LL is precisely equal to $N_\phi$ in the periodic geometry, the relation Eq.~\ref{fillingfactor} has no shift for small systems (in contrast to the spherical geometry).

$\Psi_n$ is a determinant composed of the appropriate single-particle states. 
It is convenient to express the wave function as 
\begin{equation}
\label{CF general form}
\Psi^{\rm unproj}_\frac{n}{2pn+1}[z_i,\bar{z_i}]=e^{\frac{\sum_i (z_i^2-|z_i|^2)}{4l^2}}\chi_n[{f}_i(z_j)] (\chi_1[{f}_i(z_j)])^{2p}
\ee
where
\be
\label{upro chi}
\chi_n[{f}_i(z_j)]=
\begin{vmatrix}
f_1^{(1)}(z_1)&f_1^{(1)}(z_2)&\ldots&f_1^{(1)}(z_N) \\
f_1^{(2)}(z_1)&f_1^{(2)}(z_2)&\ldots &f_1^{(2)}(z_N)\\
\vdots&\vdots&\vdots \\
f_1^{(N_\phi)}(z_1)&f_1^{(N_\phi)}(z_2)&\ldots&f_1^{(N_\phi)}(z_N) \\
f_2^{(1)}(z_1,\bar{z}_1)&f_2^{(1)}(z_2,\bar{z}_2)&\ldots&f_2^{(1)}(z_N,\bar{z}_N) \\
f_2^{(2)}(z_1,\bar{z}_1)&f_2^{(2)}(z_2,\bar{z}_2)&\ldots &f_2^{(2)}(z_N,\bar{z}_N)\\
\vdots&\vdots&\vdots \\
f_{n}^{(N_\phi)}(z_1,\bar{z}_1)&f_{n}^{(N_\phi)}(z_2,\bar{z}_2)&\ldots&f_{n}^{(N_\phi)}(z_N,\bar{z}_N)\\
\end{vmatrix}
\end{equation}
The single particle wave functions $f_{n}^{(i)}(z_1,\bar{z}_1)$ were given in Eq.~\ref{higher f}.

The wave function $\Psi$ satisfies the periodic boundary conditions given in Eq.~\ref{pb} provided that the single-particle wave functions in $\chi_n$ and $\chi_1$ satisfy Eq.~\ref{pb},
and the various phases satisfy
\be
\label{pbs for product}
\begin{gathered}
\phi_1=\phi_1^{(n)}+2p\phi_1^{(1)}\\
\phi_\tau=\phi_\tau^{(n)}+2p\phi_\tau^{(1)}
\end{gathered}
\ee

We note that the wave function in Eq.~\ref{CF product} does not, in general, have a well-defined CM momentum. To see this, we recall that the CM momentum is defined by the eigenvalue of the \rm{CM} magnetic translation operator 
\be
t_{\rm \rm{CM}}\left(\frac{L_1}{N_\phi}\right)=\prod_{i=1}^{N}t\left(\frac{L_1}{N_\phi}\right)
\ee
where $L_1/N_\phi$ is the smallest discrete value that preserves the boundary conditions \cite{Bernevig12,Haldane85b}.
While $\Psi_n$ is the eigenstate of $t_{\rm{CM}}\left(\frac{L_1}{N^*_\phi}\right)$ and $\Psi_1$ is the eigenstate of $t_{\rm \rm{CM}}\left(\frac{L_1}{N}\right)$, the product $\Psi^{\rm unproj}_{n\over 2pn+1}$ is not an eigenstate of  $t_{\rm \rm{CM}}\left(\frac{L_1}{N_\phi}\right)$, since $\frac{L_1}{N_\phi}$ is smaller than both $\frac{L_1}{N^*_\phi}$ and $\frac{L_1}{N}$.

It is known from general considerations \cite{Haldane85,Haldane85b,Bernevig12,Greiter16} that the ground state at $\nu=n/(2pn+1)$ has a degeneracy of $2pn+1$, with the different ground states related by the CM magnetic translation. The wave function for ground state at $\nu=\frac{n}{2pn+1}$ obtained above is thus a superposition of the $2pn+1$ CM eigenstates. That is not a problem for the calculation of many observable quantities, such as the energy or the pair correlation function, because they do not depend on the CM part of the wave function. Nonetheless, it would be important to derive explicitly the correct degeneracy of the ground state. For Laughlin's wave function at $\nu=1/(2p+1)$, the CM part factors out which allows an explicit construction of the $2p+1$ wave functions that are eigenstates of the CM operator, as shown in Appendix \ref{CMLaughlin}.  In Appendix~\ref{CM deg} we show how, starting from the wave function in Eq.~\ref{CF general form}, we can construct $2pn+1$ degenerate states at $\nu=n/(2pn+1)$ with well-defined CM momenta.

\subsection{LLL projection of products of single particle wave functions}

In the CF theory, we need to project the products of wave functions to the LLL. An advantage of using the symmetric gauge is that in the torus geometry the projection method is analogous to that in the disk geometry \cite{Jain97,Jain97b,Jain07}. However, we need to check that the projected wave functions satisfy the correct periodic boundary conditions for individual particles. 

In this section, we prove the following result:
\be
P_{\rm LLL} \psi_n \psi'_1 \psi_1''=e^{z^2-|z|^2\over 4l^2} \hat{f}_n f_1' f_1''
\ee
where $\hat{f}_n$ is an operator that does not act on the Gaussian and exponential factors (which have been moved to the left) and does not depend on the wave function (e.g., $f_1' f_1''$) on which it is acting (provided it is in the LLL). Following the standard method of LLL projection, we have 
\be
e^{z^2-|z|^2\over 4l^2}\hat{f}_n=e^{-|z|^2\over 4l^2} f_n(\bar{z}\rightarrow 2l^2\partial/\partial z, z) e^{z^2\over 4l^2}
\ee
It should be understood here and below that in $f_n(\bar{z}\rightarrow 2l^2\partial/\partial z, z)$, $\bar{z}$ is moved to the far left before making the replacement $\bar{z}\rightarrow 2l^2\partial/\partial z$. 

Let us illustrate how to derive $\hat{f}_n$ by taking an example in the second LL. 
First, we write out the unprojected wave function with the Gaussian factor $e^{-\frac{|z|^2}{4l^2}}$ on the far left:
\begin{widetext}
\ba
\label{prod LLL pro}
\psi(z,\bar{z})&=&\psi_2(z,\bar{z}) \psi_1'(z,\bar{z}) \psi_1''(z,\bar{z}) \nonumber \\
&=&e^{\frac{-|z|^2}{4l^2}} \left[\frac{\bar{z}}{2l^{*2}}e^{\frac{z^2}{4l^2}}f_1(z)f_1'(z) f_1''(z)-\left(\frac{z}{2l^{*2}}f_1(z)+\partial_z f_1(z)\right)e^{\frac{z^2}{4l^2}}f_1'(z) f_1''(z)\right]
\ea
\end{widetext}
Here $l$ is the physical magnetic length and $l^*$ is the effective magnetic length for composite fermions satisfying
\be
\label{l*}
\frac{l^{*2}}{l^2}=\frac{N_\phi}{|N_\phi^*|}
\ee
Next, we replace $\bar{z}$ with $2l^2\partial_z$ and let it act on the rest of the wave function:
\ba
\label{LLL single}
&&P_{\rm LLL}\psi_2(z,\bar{z}) \psi_1'(z,\bar{z}) \psi_1''(z,\bar{z}) \nonumber\\
&=& e^{\frac{z^2-|z|^2}{4l^2}} 
\left[
\frac{l^2-l^{*2}}{l^{*2}}
\frac{\partial f_1}{\partial z}  f_1' f_1''+\frac{l^2}{l^{*2}}f_1\partial_z (f_1'f_1'')
\right] 
\nonumber \\
&=& e^{z^2-|z|^2\over 4l^2} \hat{f}_2 f_1' f_1''\nonumber \\
&\equiv&e^{z^2-|z|^2\over 4l^2} f(z)
\ea
with 
\be
\label{2nd single}
\hat{f}_2^{(n)}(z)=\frac{l^2-l^{*2}}{l^{*2}}\frac{\partial f_1^{(n)}}{\partial z}+\frac{l^2}{l^{*2}}f_1^{(n)}{\partial\over \partial z}
\ee
where we have now restored the momentum index $n$. (We shall often suppress the dependence on $\bar{z}$ or $\partial/\partial z$ to avoid clutter.) The important point is that the form of $\hat{f}_2^{(n)}(z)$ does not depend on the wave function on which it acts, so long as the wave function is in the LLL.

We need to check whether Eq.~\ref{LLL single} satisfies the correct periodic boundary conditions. From the periodic boundary conditions on the product of unprojected single-particle wave functions, we know the following:
\be
\begin{gathered}
N_\phi^{(1)}+N_\phi^{(2)}+N_\phi^{(3)}=N_\phi\\
\phi_1^{(1)}+\phi_1^{(2)}+\phi_1^{(3)}=\phi_1\\
\phi_\tau^{(1)}+\phi_\tau^{(2)}+\phi_\tau^{(3)}=\phi_\tau\\
\end{gathered}
\ee
From algebra it follows that $f(z)$ defined in Eq.~\ref{LLL single} satisfies the first equation of Eq.~\ref{pb f(z) 1}:
 \be
 \frac{T(L_1)f(z)}{f(z)}=e^{i\phi_1}
 \ee
To check that $f(z)$ also satisfies the second equation of Eq.~\ref{pb f(z) 1}, let us apply $T(L_1\tau)$ on $f(z)$:
 \begin{widetext}
 \ba
 T(L_1\tau)f(z)&=&\frac{l^2-l^{*2}}{l^{*2}}\left[-\frac{i2\pi}{L_1}N_\phi^{(1)}e^{i\left(\phi_\tau-\pi N_\phi\left(\frac{2z}{L_1}+\tau\right)\right)}f_1f'_1f''_1+e^{i\left(\phi_\tau-\pi N_\phi\left(\frac{2z}{L_1}+\tau\right)\right)}\frac{\partial f_1}{\partial z}f'_1f''_1\right]\nonumber \\
 &+&\frac{l^2}{l^{*2}}\left[-\frac{i2\pi}{L_1}\left(N_\phi^{(2)}+N_\phi^{(3)}\right)e^{i\left(\phi_\tau-\pi N_\phi \left(\frac{2z}{L_1}+\tau \right)\right)}f_1f'_1f''_1+e^{i\left(\phi_\tau-\pi N_\phi \left(\frac{2z}{L_1}+\tau \right) \right)}f_1\partial_z \left(f'_1f''_1\right)\right]
\ea
\end{widetext}
The first terms inside both sets of large square brackets cancel because
\be
(l^{*2}-l^2)N_\phi^{(1)}=l^2(N_\phi^{(2)}+N_\phi^{(3)})
\ee
Then we have
\be
 T(L_1\tau)f(z)=e^{i(\phi_\tau-\pi N_\phi(\frac{2z}{L_1}+\tau))}f(z)
 \ee
 The periodic boundary conditions are therefore indeed preserved. Of course, that is expected from the fact that the unprojected product wave function satisfies the correct periodic boundary conditions, and because its LLL and higher LL components are orthogonal, they must both separately satisfy the correct periodic boundary conditions. 
 
Similarly, it can be shown that the operator corresponding to a single-particle wave function in the third $\Lambda$L is (with the momentum index $n$)
\begin{widetext}
\be
\label{3rd single}
\hat{f}_3^{(n)}(z)=\frac{l^{*2}-l^2}{2l^{*4}}f_1^{(n)}(z)+\frac{(l^{*2}-l^2)^2}{l^{*4}}\frac{\partial^2 f_1^{(n)}(z)}{\partial z^2}+\frac{2l^2(l^2-l^{*2})}{l^{*4}}\frac{\partial  f_1^{(n)}(z)}{\partial z}\frac{\partial}{\partial z}+\frac{l^4}{l^{*4}}f_1^{(n)}(z)\frac{\partial^2}{\partial z^2}
\ee
\end{widetext}

\subsection{Direct projection}

Using the results from the previous section, the LLL projected wave function can be written as
\begin{widetext}
\ba
\label{ExactPROJ}
\Psi^{\rm Direct}_{\nu^*\over 2p\nu^*+1}[z_i,\bar{z_i}]&=&e^{\frac{\sum_i (z_i^2-|z_i|^2)}{4l^2}}{\chi}_{\nu^*}[\hat{f}_i(\partial/\partial z_j,z_j)]\chi_1^{2p} \nonumber \\
&=&e^{\frac{\sum_i (z_i^2-|z_i|^2)}{4l^2}}{\chi}_{\nu^*}[\hat{f}_i(\partial/\partial z_j,z_j)]F_1^{2p}(Z)\prod_{i<j}^N \left[\theta \left(\frac{z_i-z_j}{L_1}|\tau \right)\right]^{2p}
\ea
\end{widetext}

Even though Eq.~\ref{ExactPROJ} gives a LLL projected wave function with correct periodic boundary conditions, it is not possible to explicitly evaluate it except for small systems. The reason is that the projection requires keeping track of all Slater determinant basis functions, the number of which grows exponentially with the number of particles, $N$. This problem was circumvented in the disk and spherical geometries through another projection method, called the JK projection, to which we now come.

\subsection{JK projection: review for disk geometry}
\label{JKreview}

Let us briefly review the JK projection for the disk geometry. The notation in this subsection will be slightly different from that in the rest of the paper, but should be self-explanatory. 

The unprojected wave functions in the disk geometry have the form 
\be
\Psi^{\rm unproj}_{\nu^*\over 2\nu^*+1}=e^{-\sum_j {|z_j|^2\over 4l^2}} \chi_{\nu^*}\left(f_i(\bar{z}_j,z_j)\right) \prod_{j<k}(z_j-z_k)^{2p}
\ee
where $f_i(\bar{z}_j,z_j)$ are single particle wave functions, with $i$ collectively denoting the LL and momentum quantum numbers. The Direct projection is obtained as 
\be
\Psi_{\nu^*\over 2\nu^*+1}=e^{-\sum_j {|z_j|^2\over 4l^2}} \chi_{\nu^*}\left(\hat{f}_i(\partial/\partial z_j,z_j)\right) \prod_{j<k}(z_j-z_k)^{2p}
\ee
where $\hat{f}_i(\partial/\partial z_j,z_j)=f_i(\bar{z}_j\rightarrow 2l^2\partial/\partial z_j,z_j)$. As discussed above, it is not possible to evaluate this wave function for large $N$. To make further progress, we write, following JK:
 \be
\Psi^{\rm JK}_{\nu^*\over 2\nu^*+1}=e^{-\sum_j {|z_j|^2\over 4l^2}} \chi_{\nu^*}\left(\hat{f}_i(\partial/\partial z_j,z_j)J_j^p\right) 
\ee
where 
\be
J_i=\prod_{j(j\neq i)}\left(z_i-z_j\right)
\ee
In the JK wave function, one projects each term of the Slater determinant $\chi$ individually. One thus needs to evaluate a single Slater determinant for the FQH ground and excited states, which enables a study of very large systems.

\subsection{Failure of JK projection for the torus geometry}
\label{Failure of the JK projection method}

In this section, we show that if we directly apply the JK projection method as it is implemented in disk and spherical geometries, it does not produce a valid wave function in the torus geometry, because the resulting wave function does not satisfy the correct periodic boundary conditions. 

Seeking to generalize the JK projection to the torus geometry, we note that the factor $\prod_{i<j}^N \left[\theta \left(\frac{z_i-z_j}{L_1}|\tau \right)\right]^{2p}$ is quite analogous to the Jastrow factor of the disk geometry, but the presence of the CM factor $F_1^2(Z)$ seems to pose a difficulty. 
Fortunately, as shown in Appendix \ref{proof that operator commute with F(Z)}, the operator ${\chi}_{\nu^*}[\hat{f}_i(\partial/\partial z_j,z_j)]$ commutes with $F_1^2(Z)$ for all proper states. We can thus incorporate the Jastrow factor $\prod_{i<j}^N \left[\theta \left(\frac{z_i-z_j}{L_1}|\tau \right)\right]^{2p}$ into ${\chi}_{\nu^*}[\hat{f}_i(\partial/\partial z_j,z_j)]$  as follows:
\be
\label{projected wf2}
\Psi^{\rm JK}_\frac{\nu^*}{2p\nu^*+1}[z_i,\bar{z_i}]=e^{\frac{\sum_i (z_i^2-|z_i|^2)}{4l^2}}F_1^2(Z){\chi}^{\rm JK}_{\nu^*}[\hat{f}_i(\partial/\partial z_j,z_j)J^p_j]
\ee
with
\be
J_i=\prod_{j(j\neq i)}\theta \left(\frac{z_i-z_j}{L_1}|\tau \right)\\
\ee
This is not a valid wave function, however. To show that it violates the periodic boundary conditions, we take the $\nu=2/5$ state with $N=2,N_\phi=5$ as an example. In this case, we can write the determinant explicitly (note that there is only one eigenstate in each Landau level, so we suppress the superscript):
\begin{widetext}
\be
\label{2 5}
{\chi}^{\rm JK}_2[\hat{f}_i(\partial/\partial z_j,z_j)J_j]=
\begin{vmatrix}
{f_1}(z_1)\theta \left(\frac{z_1-z_2}{L_1}|\tau \right)&{f_1}(z_2)\theta \left(\frac{z_2-z_1}{L_1}|\tau \right)\\
-\frac{4}{5}\frac{\partial f_1(z_1)}{\partial z_1}\theta \left(\frac{z_1-z_2}{L_1}|\tau \right)+\frac{1}{5}f_1(z_1)\frac{\partial \theta \left(\frac{z_1-z_2}{L_1}|\tau \right)}{\partial z_1}&-\frac{4}{5}\frac{\partial f_1(z_2)}{\partial z_2}\theta \left(\frac{z_2-z_1}{L_1}|\tau \right)+\frac{1}{5}f_1(z_2)\frac{\partial \theta \left(\frac{z_2-z_1}{L_1}|\tau \right)}{\partial z_2})\\
\end{vmatrix}
\ee
\end{widetext}
Here we have 
\be
 \label{ground 2/5}
 \hat{f}_2^{(n)}(z,\bar{z})=-\frac{4}{5}\frac{\partial f_1^{(n)}(z)}{\partial z}+\frac{1}{5}f_1^{(n)}(z)\frac{\partial}{\partial z}
 \ee
which follows from Eq.~\ref{2nd single} noting that at filling factor $\nu=2/5$, we have $\frac{l^{*2}}{l^2}=5$.
 
To satisfy the periodic boundary condition in the $L_1 \tau$ direction, ${\chi}^{\rm JK}_2$ needs to satisfy (for a translation of the particle 1)
\be
T_1(L_1\tau){\chi}^{\rm JK}_2=e^{i(2\pi (2Z/L_1+\tau)-\pi N_\phi(2z_1/L_1+\tau))}{\chi}^{\rm JK}_2
\ee
However, an explicit calculation gives
\begin{widetext}
\begin{multline}
T_1(L_1\tau){\chi}^{\rm JK}_2=e^{i(2\pi (2Z/L_1+\tau)-\pi N_\phi(2z_1/L_1+\tau))}{\chi}^{\rm JK}_2+\\ \frac{1}{5}\frac{i2\pi}{L_1}e^{i(2\pi (2Z/L_1+\tau)-\pi N_\phi(2z_1/L_1+\tau))}f_1(z_1)f_1(z_2)\theta \left(\frac{z_1-z_2}{L_1}|\tau \right)\theta \left(\frac{z_2-z_1}{L_1}|\tau \right)
\end{multline}
\end{widetext}
indicating that the wave function does not satisfy the periodic boundary conditions. This may seem to make the JK projection method unimplementable in the torus geometry, which would make it impractical to do calculations with the CF theory in the torus geometry. However, we show in the next section that, fortunately, it is possible to modify the JK projection method to obtain legitimate LLL wave functions.

\subsection{Modified LLL projection method}
\label{Modified LLL projection method}

The two particle problem considered in the previous section gives us a clue that leads us to an elegant solution for how the JK projection method can be modified to produce legitimate wave functions. Let us first go back to the direct projection of the system:
\be
\Psi^{\rm Direct}_\frac{2}{5}[z_i,\bar{z_i}]=e^{\frac{\sum_i (z_i^2-|z_i|^2)}{4l^2}}F_1^2(Z){\chi}_2(\hat{f}_i(\partial/\partial z_j,z_j)) J^2
\ee
where $J= \theta \left(\frac{z_1-z_2}{L_1}|\tau \right)$.
This of course satisfies the correct periodic boundary conditions. 
The factor ${\chi}_2[\hat{f}_i(\partial/\partial z_j,z_j)]J^2$ can be written as
\begin{widetext}
\ba
\label{2 5 2}
{\chi}_2[\hat{f}_i(\partial/\partial z_j,z_j)]J^2&=&
\begin{vmatrix}
{f_1}(z_1)&{f_1}(z_2)\\
-\frac{4}{5}\frac{\partial f_1(z_1)}{\partial z_1}+\frac{1}{5}f_1(z_1)\frac{\partial}{\partial z_1}&-\frac{4}{5}\frac{\partial f_1(z_2)}{\partial z_2}+\frac{1}{5}f_1(z_2)\frac{\partial}{\partial z_2}\\
\end{vmatrix}
J^2 \nonumber\\
&=&
\begin{vmatrix}
{f_1}(z_1)&{f_1}(z_2)\\
-\frac{4}{5}\frac{\partial f_1(z_1)}{\partial z_1}J^2+\frac{1}{5}f_1(z_1)\frac{\partial J^2}{\partial z_1}&-\frac{4}{5}\frac{\partial f_1(z_2)}{\partial z_2}J^2+\frac{1}{5}f_1(z_2)\frac{\partial J^2}{\partial z_2} \nonumber\\
\end{vmatrix}\\
&=&
\begin{vmatrix}
{f_1}(z_1)J&{f_1}(z_2)J\\
-\frac{4}{5}\frac{\partial f_1(z_1)}{\partial z_1}J+\frac{2}{5}f_1(z_1)\frac{\partial J}{\partial z_1}&-\frac{4}{5}\frac{\partial f_1(z_2)}{\partial z_2}J+\frac{2}{5}f_1(z_2)\frac{\partial J}{\partial z_2}\\
\end{vmatrix}
\ea
\end{widetext}
We notice that this form is almost the same as that in Eq.~\ref{2 5}, except that the coefficient of $f_1(z_i)\frac{\partial J}{\partial z_i}$ is $2 \over 5$ instead of $1 \over 5$. (The reader may notice that the second columns of Eq.~\ref{2 5} and Eq.~\ref{2 5 2} have opposite signs, but that merely contributes an unimportant $-1$ to the overall normalization factor.)

This suggests a possible way to modify the JK projection. We ask whether replacing $\hat{f}_n^{(m)}$ by a related operator $\hat{g}_n^{(m)}$ could give a wave function with the correct boundary conditions. Let us specialize to the second $\Lambda$L and try the form for $\hat{g}_2^{(m)}$: 
\be
\label{unk g}
\hat{g}_2^{(n)}(z)=-\frac{N_\phi-N_\phi^*}{N_\phi}\frac{\partial f_1^{(n)}(z)}{\partial z}+\alpha\frac{N_\phi^*}{N_\phi}f_1^{(n)}(z)\frac{\partial}{\partial z}
\ee
where $\alpha$ is an unknown coefficient. We now ask whether a value for $\alpha$ can be found that produces a wave function that satisfies correct boundary conditions.

We consider a general wave function of the type
 \be
 \label{projected wf}
\psi[z_i,\bar{z_i}]=e^{\frac{\sum_i (z_i^2-|z_i|^2)}{4l^2}}F_1^{2p}(Z){\chi}[\hat{g}_i(z_j)J_j^p]
\ee
\be
\label{chi-det}
{\chi}[\hat{g}_i(z_j)J_j^p]=
\begin{vmatrix}
\hat{g_1}^{(1)}(z_1)J_1^p&\ldots&\hat{g_1}^{(1)}(z_N)J_N^p \\
\vdots&\vdots&\vdots \\
\hat{g_2}^{(1)}(z_1)J_1^p&\ldots&\hat{g_2}^{(1)}(z_N)J_N^p \\
\vdots&\vdots&\vdots \\
%\hat{g_3}^{(1)}(z_1)J_1&\ldots&\hat{g_3}^{(1)}(z_N)J_N \\
%\vdots&\vdots&\vdots \\
\end{vmatrix}
\ee
where we assume that in $\chi$ the LLL is fully occupied, the second LL is arbitrarily occupied, and third and higher LLs are unoccupied. This includes the 2/5 ground state ($\chi$ has second LL fully occupied), a CF quasiparticle of the 1/3 state ($\chi$ has only a single electron in the second LL), a CF quasihole of 2/5 ($\chi$ has a single hole in the second LL), and quasidegenerate ground states at arbitrary fillings in the range $2/5\geq \nu \geq 1/3$. 

The wave function in Eq.~\ref{projected wf} should satisfy the periodic boundary conditions:
\be
\begin{gathered}
\frac{T_i(L_1)F_1^{2p}(Z){\chi}[\hat{g}_i(z_j)J_j^p]}{F_1^{2p}(Z){\chi}[\hat{g}_i(z_j)J_j^p]}=e^{i\phi_1}\\
\frac{T_i(L_1\tau)F_1^{2p}(Z){\chi}[\hat{g}_i(z_j)J_j^p]}{F_1^{2p}(Z){\chi}[\hat{g}_i(z_j)J_j^p]}=e^{i(\phi_\tau-\pi N_\phi(2z_i/L_1+\tau))}\\
\end{gathered}
\ee
For convenience, we will take $\phi_1=0,\phi_\tau=0$. Considering the periodic properties of $F_1^2(Z)$, the periodic boundary conditions for ${\chi}[\hat{g}_i(z_j)]$ are:
\be
\label{chi pb}
\begin{gathered}
\frac{T_i(L_1){\chi}[\hat{g}_i(z_j)J_j^p]}{{\chi}[\hat{g}_i(z_j)J_j^p]}=1\\
\frac{T_i(L_1\tau){\chi}[\hat{g}_i(z_j)J_j^p]}{{\chi}[\hat{g}_i(z_j)J_j^p]}=e^{i(2p\pi (2Z/L_1+\tau)-\pi N_\phi(2z_i/L_1+\tau))}\\
\end{gathered}
\ee
Explicit calculation shows that the first equation in Eq.~\ref{chi pb} is automatically satisfied for $\hat{g}_2^{(m)}$ with any value of $\alpha$. The key is the second equation in Eq.~\ref{chi pb}. 

Translating $z_i$ by $L_1\tau$ gives
\begin{widetext}
\be
T_i(L_1\tau)\hat{g}_1^{(n)}(z_j)J_j^p=e^{ip\pi \left(\frac{2(z_j-z_i)}{L_1}-\tau+1\right)}\hat{g}_1^{(n)}(z_j)J_j^p,j\neq i
\label{L1tau1}
\ee
\be
T_i(L_1\tau)\hat{g}_1^{(n)}(z_i)J_i^p=e^{-i\pi N_\phi^*\left(\frac{2z_i}{L_1}+\tau\right)}\prod_{j(j\neq i)}e^{-ip\pi \left(\frac{2(z_i-z_j)}{L_1}+\tau+1\right)}\hat{g}_1^{(n)}(z_i)J_i^p
\label{L1tau2}
\ee
\be
\label{L1tau3}
T_i(L_1\tau)\hat{g}_2^{(n)}(z_j)J_j^p=e^{ip\pi \left(\frac{2(z_j-z_i)}{L_1}-\tau+1\right)}\hat{g}_2^{(n)}(z_j)J_j^p+{\color{blue}[p\alpha]}\frac{i2\pi N_\phi^*}{L_1N_\phi} e^{ip\pi \left(\frac{2(z_j-z_i)}{L_1}-\tau+1\right)}\hat{g}_1^{(n)}(z_j)J_j^p,j\neq i
\ee
\ba
T_i(L_1\tau)\hat{g}_2^{(n)}(z_i)J_i^p &=& e^{-i\pi N_\phi^*\left(\frac{2z_i}{L_1}+\tau \right)}\prod_{j(j\neq i)}e^{-ip\pi \left(\frac{2(z_i-z_j)}{L_1}+\tau+1\right)}\hat{g}_2^{(n)}(z_i)J_i^p
\nonumber
\\&&+{\color{blue}\left[N_\phi-N_\phi^*-p\alpha N+p\alpha \right]}\frac{i2\pi N_\phi^*}{L_1N_\phi}e^{-i\pi N_\phi^*\left(\frac{2z_i}{L_1}+\tau \right)}\prod_{j(j\neq i)}e^{-ip\pi \left(\frac{2(z_i-z_j)}{L_1}+\tau+1\right)}\hat{g}_1^{(n)}(z_i)J_i^p
\label{L1tau4}
\ea
\end{widetext}
It is the second terms in Eqs.~\ref{L1tau3} and \ref{L1tau4} that violate the periodic boundary conditions.  Without those terms, it can be seen, by taking the product of the factors from each column, that ${\chi}[\hat{g}_i(z_j)J_j^p]$ would satisfy the correct boundary conditions in Eq.~\ref{chi pb}. It turns out that the second terms are eliminated if we choose the blue-highlighted terms in the square brackets ${\color{blue}[\ldots]}$ in Eqs.~\ref{L1tau3} and \ref{L1tau4} to be equal:
\be
\label{alpha}
p\alpha =N_\phi-N_\phi^*-p\alpha N+p\alpha 
\ee
which, with $N_\phi=N_\phi^*+2pN$, reduces to $\alpha=2$. With this choice, the second terms in Eqs.~\ref{L1tau3} and \ref{L1tau4} are expunged from the Slater determinant $\chi$ of Eq.~\ref{chi-det}, because they are proportional to the corresponding row in the LLL containing the terms given in Eqs.~\ref{L1tau1} and ~\ref{L1tau2}.

Thus we have
 \be
 \label{2nd LL matrix element}
 \hat{g}_2^{(n)}(z)=-\frac{N_\phi-N_\phi^*}{N_\phi}\frac{\partial f_1^{(n)}(z)}{\partial z}+\frac{N_\phi^*}{N_\phi}f_1^{(n)}(z){\color{red} {\bf 2}}\frac{\partial}{\partial z}
 \ee
A similar but more lengthy algebra (which we leave out) shows that for the third LL, the choice 
\begin{widetext}
\be
\label{3rd LL matrix element}
\hat{g}_3^{(n)}(z)=\frac{N_\phi-N_\phi^*}{2N_\phi^2}f_1^{(n)}(z)+\frac{(N_\phi-N_\phi^*)^2}{N_\phi^2}\frac{\partial^2 f_1^{(n)}(z)}{\partial z^2}-\frac{2N_\phi^*(N_\phi-N_\phi^*)}{N_\phi^2}\frac{\partial  f_1^{(n)}(z)}{\partial z}{\color{red} {\bf 2}}\frac{\partial}{\partial z}+\frac{N_\phi^{*2}}{N_\phi^2}f_1^{(n)}(z)\left({\color{red} {\bf 2}}\frac{\partial}{\partial z}\right)^2
\ee
\end{widetext}
produces wave functions with the correct boundary conditions, provided that the lowest two $\Lambda$Ls are fully occupied. The operators $\hat{g}$ in Eqs.~\ref{2nd LL matrix element} and \ref{3rd LL matrix element} differ from $\hat{f}$ in Eqs.~\ref{2nd single} and \ref{3rd single} only through the factors highlighted in red. 

One may ask whether $\hat{g}_n^{(m)}(z_j)$ exists for yet higher LLs. The answer is in the affirmative. The derivation for $\hat{g}_n^{(m)}(z_j)$ for arbitrary LL is given in Appendix \ref{proof modified JK}. Interestingly, the general rule is that we can go from $\hat{f}$ to $\hat{g}$ by making the replacement $\partial/\partial z \rightarrow 2\partial/\partial z$  for the derivatives acting on $J_i$; Eqs.~\ref{2nd LL matrix element} and \ref{3rd LL matrix element} for $\hat{g}_2^{(m)}(z_j)$ and $\hat{g}_3^{(m)}(z_j)$ are written so as to make this explicit. 

The crucial aspect that renders the wave functions in Eq.~\ref{projected wf} valid is that the unwanted terms in each row are eliminated by the rows corresponding to single-particle states in lower levels with the same momentum quantum numbers. This implies that the modified projection method produces valid wave functions for all proper states defined in the introduction.

\section{Testing the accuracy of the LLL projected sates}
\label{Testing the validity of the LLL projected sates}

In the previous section, we have shown how we can modify the JK projection method in the torus geometry to obtain LLL wave functions that satisfy the correct boundary conditions. However, there is no guarantee that they are accurate representations of the Coulomb eigenstates. That must be ascertained by a direct comparison.
In this section, we perform such comparisons for the ground states, CF quasiparticles, and CF quasiholes at $\nu=1/3$ and $\nu=2/5$. We also evaluate the pair correlation function.  The reader may refer to Appendix \ref{interaction} for the standard definition of the periodic Coulomb interaction in the torus geometry, as well as certain other technical details for our Monte Carlo calculations. In all our numerical evaluations, we choose a square torus, i.e. $\tau=i$.

The ground state energies for $\nu=1/3$, $2/5$, and $3/7$ are shown in Tables~\ref{1_3ground}, \ref{2_5ground}, and \ref{3_7ground}. The exact diagonalization energies are also given wherever available. The thermodynamic limits are shown in Fig.~\ref{25ground} (small systems not used in the extrapolation are not shown) as well as in Tables~\ref{1_3ground}-\ref{3_7ground}. 

Comparison with exact diagonalization results establishes that our wave functions are quantitatively extremely accurate. For example, for 12 particles the energies of the Jain wave functions for 2/5 and 3/7 are within 0.07\% and 0.05\%, respectively, of the corresponding exact Coulomb energies. This level of accuracy is comparable to what has been found in the spherical geometry. Furthermore, our modified wave functions can be evaluated for much larger systems than those available to exact diagonalization. We have shown results for up to 40 particles in this article, and much larger systems should be accessible with our method. 

\begin{table}[!h]
\begin{tabular}{|c|c|c|}
\hline
N&CF&Exact\\ \hline
4&$-0.41412\pm0.00004$&-0.41519\\ \hline
6&$-0.41156\pm0.00003$&-0.41190\\ \hline
8&$-0.41091\pm0.00004$&-0.41132\\ \hline
10&$-0.41058\pm0.00004$&-0.41106\\ \hline
15&$-0.41025\pm0.00004$&\\ \hline
20&$-0.41005\pm0.00005$&\\ \hline
25&$-0.40996\pm0.00005$&\\ \hline
30&$-0.40991\pm0.00005$&\\ \hline
40&$-0.40985\pm0.00003$&\\ \hline
$\infty$&$-0.40956\pm0.00002$&\\ \hline
\end{tabular}
\caption{\label{1_3ground}The Coulomb energy per particle for the ground state at $\nu=1/3$. The energy is quoted in units of $e^{2}/\epsilon l$ and includes self-interaction.  }
\end{table}

\begin{table}[!h]
\begin{tabular}{|c|c|c|}
\hline
N&CF&Exact\\ \hline
4&$-0.43992 \pm0.00002$&-0.44026\\ \hline
8&$-0.43409 \pm0.00006$&-0.43430\\ \hline
10&$-0.43376 \pm0.00007$&-0.43395\\ \hline
12&$-0.43345 \pm0.00007$&-0.43374\\ \hline
14&$-0.4333 \pm0.0001$&\\ \hline
20&$-0.43306 \pm0.00008$&\\ \hline
26&$-0.4330 \pm0.0001$&\\ \hline
30&$-0.43290 \pm0.00006$&\\ \hline
40&$-0.4328 \pm0.0001$&\\ \hline
$\infty$&$-0.43245\pm0.00004$&\\ \hline
\end{tabular}
\caption{\label{2_5ground}The Coulomb energy per particle for the ground state at $\nu=2/5$. The energy is quoted in units of $e^{2}/\epsilon l$ and includes self-interaction. }
\end{table}

\begin{table}[!h]

\begin{tabular}{|c|c|c|}
\hline
N&CF&Exact\\ \hline
3&$-0.4431\pm0.0001$&-0.4438\\ \hline
6&$-0.4436\pm0.0001$&-0.4438\\ \hline
9&$-0.4447\pm0.0001$&-0.4448\\ \hline
12&$-0.44340\pm0.00008$&-0.44360\\ \hline
15&$-0.44303\pm0.00006$&\\ \hline
21&$-0.44262\pm0.00008$&\\ \hline
24&$-0.44258\pm0.00008$&\\ \hline
30&$-0.44245\pm0.00004$&\\ \hline
39&$-0.44239\pm0.00006$&\\ \hline
$\infty$&$-0.44188\pm0.00006$&\\ \hline
\end{tabular}
\caption{\label{3_7ground}The Coulomb energy per particle for the ground state at $\nu=3/7$. The energy is quoted in units of $e^{2}/\epsilon l$ and includes self-interaction.   }
\end{table}

\begin{figure}[!h]
\begin{center}
\includegraphics[width=7.5CM,height=5.0CM]{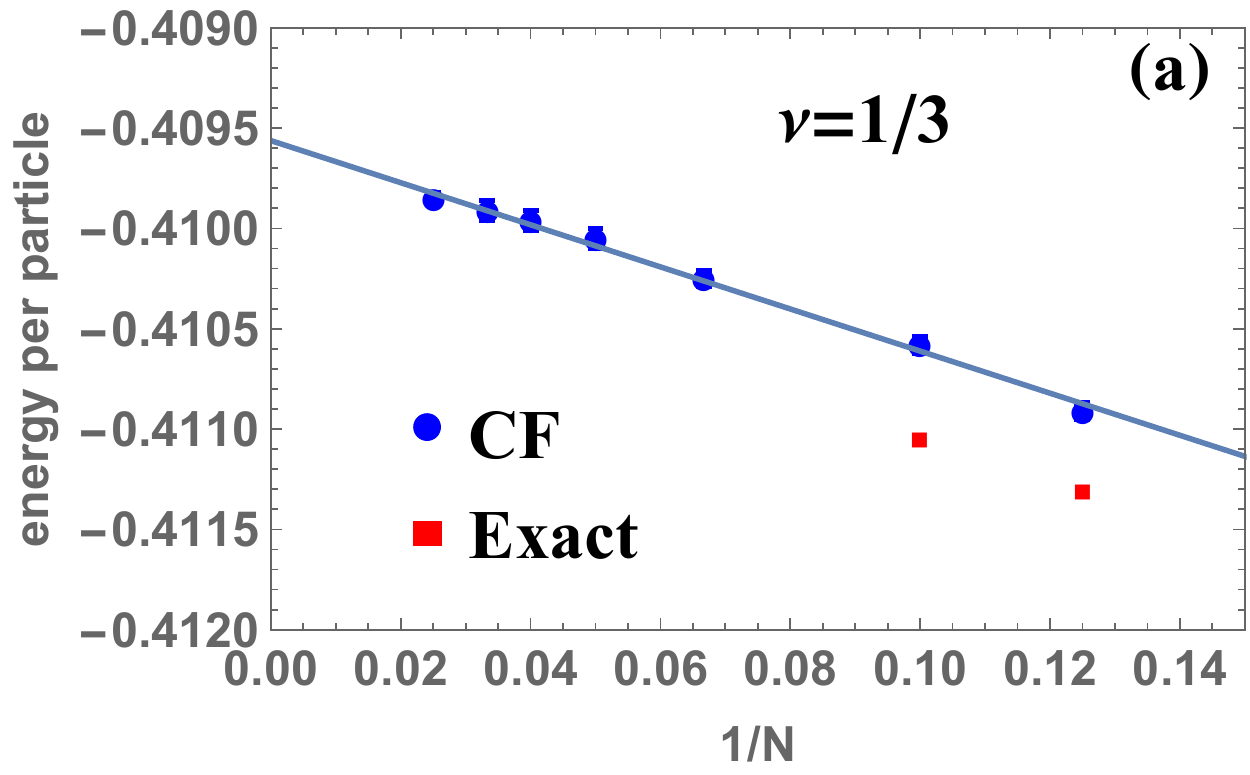}
\includegraphics[width=7.5CM,height=5.0CM]{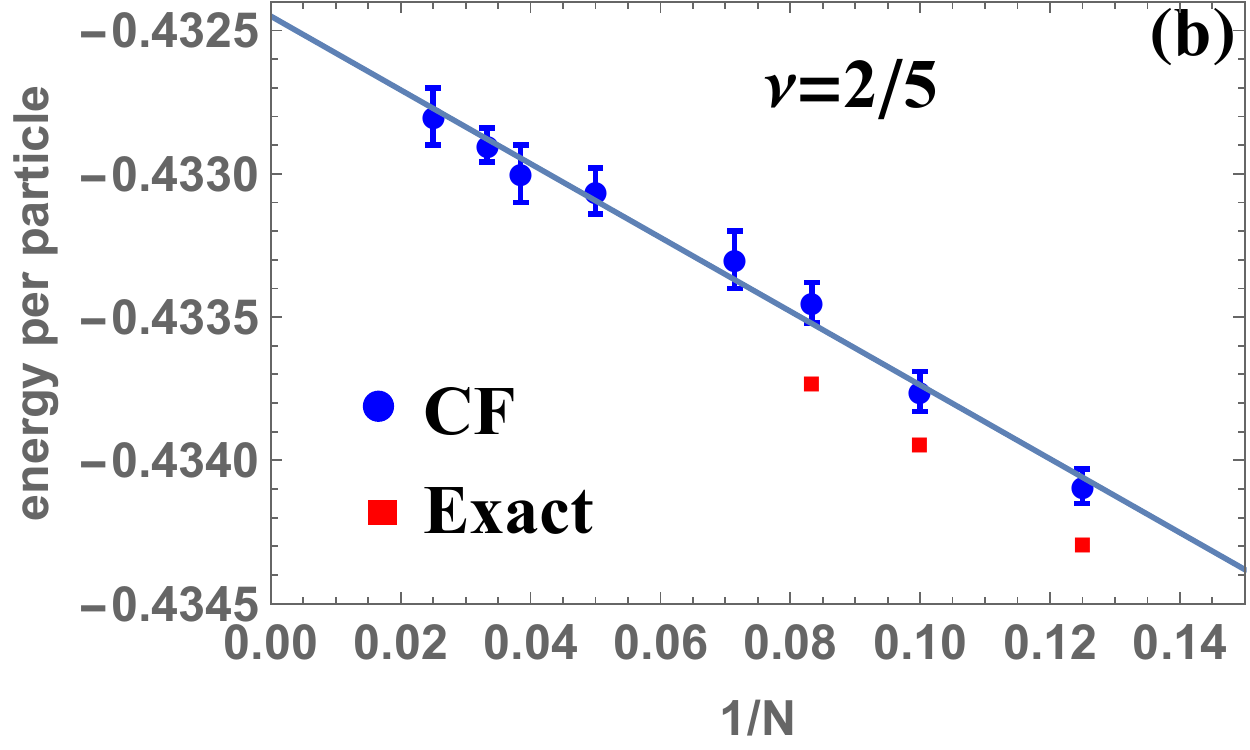}
\includegraphics[width=7.5CM,height=5.0CM]{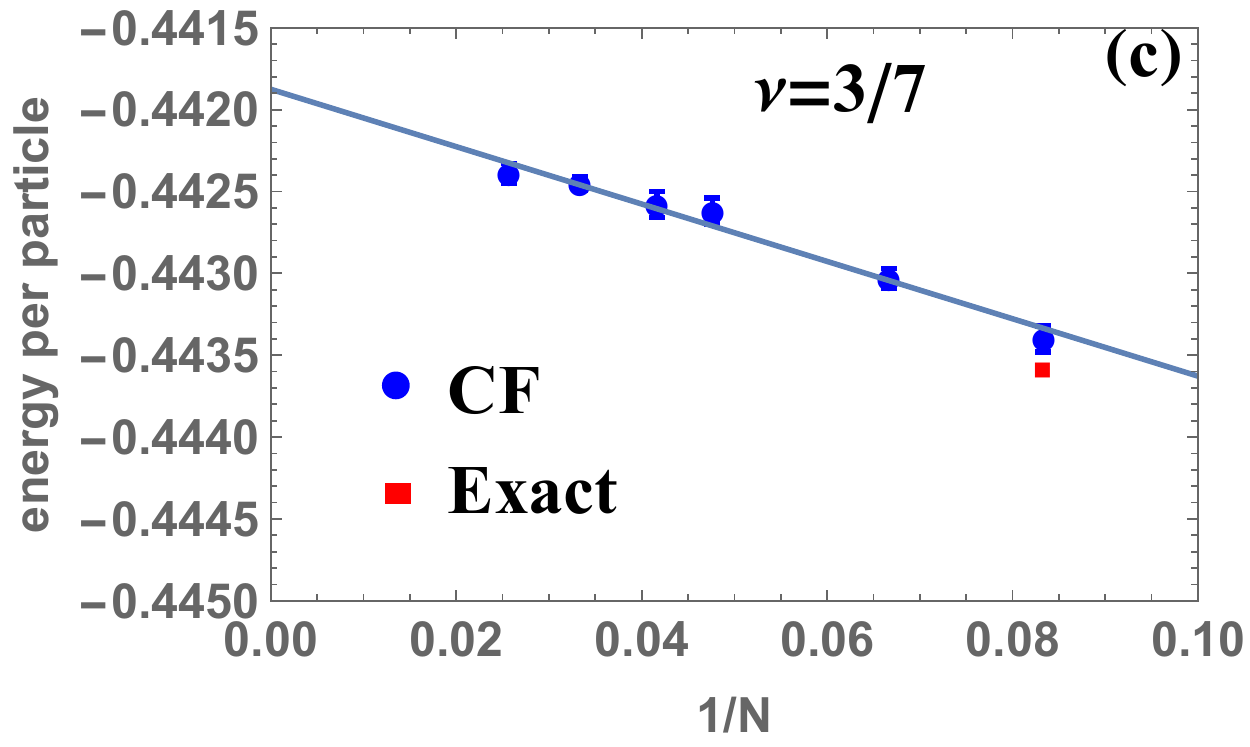} 
\end{center}
\caption{The Coulomb energy per particle for the ground states at (a) $\nu=1/3$, (b) $\nu=2/5$ and (c) $\nu=3/7$. The circles are the CF energies and the squares are the energies from exact diagonalizations. The energy is quoted in units of $e^{2}/\epsilon l$ and includes self-interaction. }
\label{25ground}
\end{figure}

An important property of a liquid state is its pair correlation function, defined as
\be
g(\vec{r})=\frac{L_1L_2}{N^2}\left<\sum_{i \neq j}\delta(\vec{r}_i-\vec{r}_j-\vec{r})\right>
\ee
It gives us the probability of finding two particles at a distance $\vec{r}$, normalized so that it approaches unity for $|\vec{r}|\rightarrow \infty$. The pair correlation function at $\nu=2/5$ for $N=30$ particles is shown in Fig.~\ref{2_5pair}. 

\begin{figure}[!h]
\begin{center}
\includegraphics[width=7.5CM,height=5.0CM]{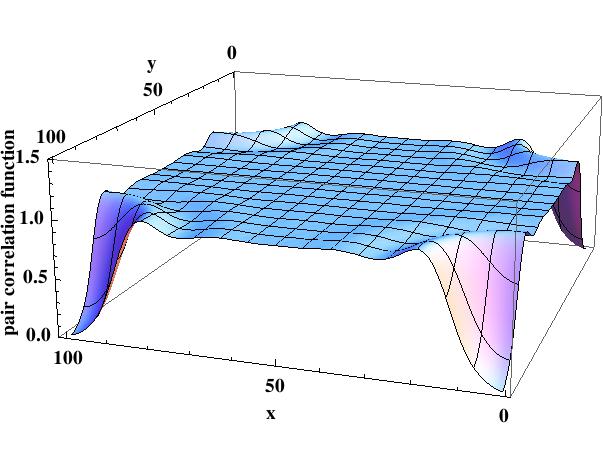} 
\end{center}
\caption{The pair correlation functions for an $N=30$ particle system at $\nu=2/5$.  }
\label{2_5pair}
\end{figure}

We have also evaluated at several filling factors the energies of the CF quasiparticle, the CF quasihole, and the excitation gap to creating a far separated CF-quasiparticle CF-quasihole pair. (Because we create the CF quasiparticle and CF quasihole separately, the sum of their energies does not include the interaction between them, and therefore corresponds to the limit of large separation.) This gap is to be identified with the activation energy measured from the Arrhenius behavior of the longitudinal resistance at low temperatures.  The CF quasihole and CF quasiparticle states for $\nu=1/3$ occur for $N_\phi=3N+1$ and $N_\phi=3N-1$, respectively. The Coulomb energies for these states are shown in Tables~\ref{1_3 hole} and ~\ref{1_3 particle}. We define the gap at $\nu=1/3$ as
\ba
\label{delta}
\Delta_{1/3}(N)&=&E^-(N,N_\phi=3N-1)+E^+(N,N_\phi=3N+1)\nonumber \\
&&-2E^0(N,N_\phi=3N)
\ea
where the first (second) term on the right-hand side is the total Coulomb energy of the $N$ particle state containing a single CF quasiparticle (CF quasihole), and $E^0(N,N_\phi=3N)$ is the energy of the $N$ particle incompressible ground state. The gaps are shown in Table~\ref{1_3 gap}. The extrapolation of the gap to the thermodynamic limit, $\frac{1}{N} \rightarrow 0$, is shown in Fig.~\ref{gapfig}(a).

At $\nu=2/5$, the incompressible ground state has an even particle number $N$, but the states containing a single CF quasiparticle or CF quasihole have an odd number of electrons. We define the gap as
\begin{multline}
\label{delta2}
\Delta_{2/5}(N)=E^-(N+1,N_\phi={5N\over 2 }+2)+\\ E^+(N-1,N_\phi={5N\over 2}-2) 
-2E^0(N,N_\phi={5N\over 2})
\end{multline}
Again, the first and second terms on the right-hand side give the total Coulomb energies of states containing a CF quasiparticle and a CF quasihole, and the last term corresponds to the ground state. All of these correspond to the same effective flux $N^*_\phi=N/2$. The total Coulomb energies for these states are shown in Tables~\ref{25par} and ~\ref{25hole} and the gaps in Table~\ref{25gap} and Fig.~\ref{gapfig}(b).

The gap energies are not as accurate as the per particle energies of the incompressible ground states, which is expected because the gaps are O(1) energies obtained by subtracting O($N$) energies. Nonetheless, the gap energies are reasonably accurate. They can be further improved, if needed, by modifying the method of CF diagonalization \cite{Mandal02} to the torus wave functions, but that is outside the scope of the current work where our goal is to demonstrate how to construct accurate wave functions for incompressible ground states, their excitations, and other proper configurations of composite fermions.

\begin{table}[!h]
\begin{tabular}{|c|c|c|}
\hline
N&CF&Exact\\ \hline
4&$-0.39552\pm0.00008$&-0.39750\\ \hline
6&$-0.39873\pm0.00003$&-0.39943\\ \hline
8&$-0.40097\pm0.00002$&-0.40170\\ \hline
10&$-0.40248\pm0.00005$&-0.40319\\ \hline
15&$-0.40467\pm0.00005$&\\ \hline
20&$-0.40575\pm0.00004$&\\ \hline
25&$-0.40648\pm0.00003$&\\ \hline
30&$-0.40699\pm0.00005$&\\ \hline
40&$-0.40761\pm0.00007$&\\ \hline
\end{tabular}
\caption{\label{1_3 hole}
The Coulomb energy per particle for $N_\phi=3N+1$, which corresponds to a single CF quasihole of the $\nu=1/3$ state. The energy is quoted in units of $e^{2}/\epsilon l$ and includes self-interaction. 
}
\end{table}

\begin{table}[!h]
\begin{tabular}{|c|c|c|}
\hline
N&CF&Exact\\ \hline
4&$-0.42050\pm0.00005$&-0.42190\\ \hline
6&$-0.41415\pm0.00007$&-0.41467\\ \hline
8&$-0.41241\pm0.00005$&-0.41303\\ \hline
10&$-0.41157\pm0.00006$&-0.41216\\ \hline
15&$-0.41068\pm0.00006$&\\ \hline
20&$-0.41024\pm0.00004$&\\ \hline
25&$-0.41005\pm0.00008$&\\ \hline
30&$-0.41000\pm0.00005$&\\ \hline
40&$-0.40988\pm0.00009$&\\ \hline
\end{tabular}
\caption{\label{1_3 particle}The Coulomb energy per particle for $N_\phi=3N-1$, which corresponds to a single CF quasiparticle of the $\nu=1/3$ state. The energy is quoted in units of $e^{2}/\epsilon l$ and includes self-interaction.}
\end{table}

\begin{table}[!h]

\begin{tabular}{|c|c|c|}
\hline
N&CF gap&Exact gap\\ \hline
4&$0.0489\pm0.0005$&0.0439\\ \hline
6&$0.0614\pm0.0006$&0.0582\\ \hline
8&$0.0675\pm0.0007$&0.0633\\ \hline
10&$0.071\pm0.001$&0.0677\\ \hline
15&$0.077\pm0.002$&\\ \hline
20&$0.082\pm0.002$&\\ \hline
25&$0.085\pm0.003$&\\ \hline
30&$0.085\pm0.004$&\\ \hline
40&$0.088\pm0.005$&\\ \hline
$\infty$&$0.095\pm0.001$&\\ \hline
\end{tabular}
\caption{\label{1_3 gap}The excitation gap for $\nu=1/3$ state in units of $e^{2}/\epsilon l$. 
}
\end{table}

\begin{table}[!h]
\begin{tabular}{|c|c|c|c|}
\hline
N&$N_\phi$&CF&Exact\\ \hline
5&12&$-0.43868\pm0.00007$&-0.43882\\ \hline
9&22&$-0.4354\pm0.0001$&-0.4361\\ \hline
11&27&$-0.43450\pm0.00006$&-0.43482\\ \hline
15&37&$-0.43393\pm0.00006$&\\ \hline
21&52&$-0.43340\pm0.00009$&\\ \hline
31&77&$-0.43307\pm0.00003$&\\ \hline
41&102&$-0.43295\pm0.00009$&\\ \hline
\end{tabular}
\caption{\label{25par}The Coulomb energy per particle for several systems containing a single CF quasiparticle of the $\nu=2/5$ state. The energy is quoted in units of $e^{2}/\epsilon l$ and includes self-interaction. }
\end{table}

\begin{table}[!h]
\begin{tabular}{|c|c|c|c|}
\hline
N&$N_\phi$&CF&Exact\\ \hline
9&23&$-0.42877\pm0.00007$&-0.42957\\ \hline
13&33&$-0.42958\pm0.00005$&\\ \hline
19&48&$-0.43043\pm0.00008$&\\ \hline
29&73&$-0.43115\pm0.00006$&\\ \hline
39&98&$-0.43154\pm0.00005$&\\ \hline
\end{tabular}
\caption{\label{25hole}The Coulomb energy per particle for several states containing a single CF quasihole of the $\nu=2/5$ state. The energy is quoted in units of $e^{2}/\epsilon l$ and includes self-interaction.}
\end{table}

\begin{table}[!h]
\begin{tabular}{|c|c|c|}
\hline
N&CF&Exact\\ \hline
10&$0.037\pm0.002$&0.030\\ \hline
14&$0.039\pm0.003$&\\ \hline
20&$0.043\pm0.004$&\\ \hline
30&$0.045\pm0.004$&\\ \hline
40&$0.043\pm0.009$&\\ \hline
$\infty$&$0.049\pm0.002$&\\ \hline
\end{tabular}
\caption{\label{25gap}The excitation gap for $\nu=\frac{2}{5}$ in units of $e^{2}/\epsilon l$. 
}
\end{table}

\begin{figure}[!h]
\begin{center}
\includegraphics[width=7.5CM,height=5.0CM]{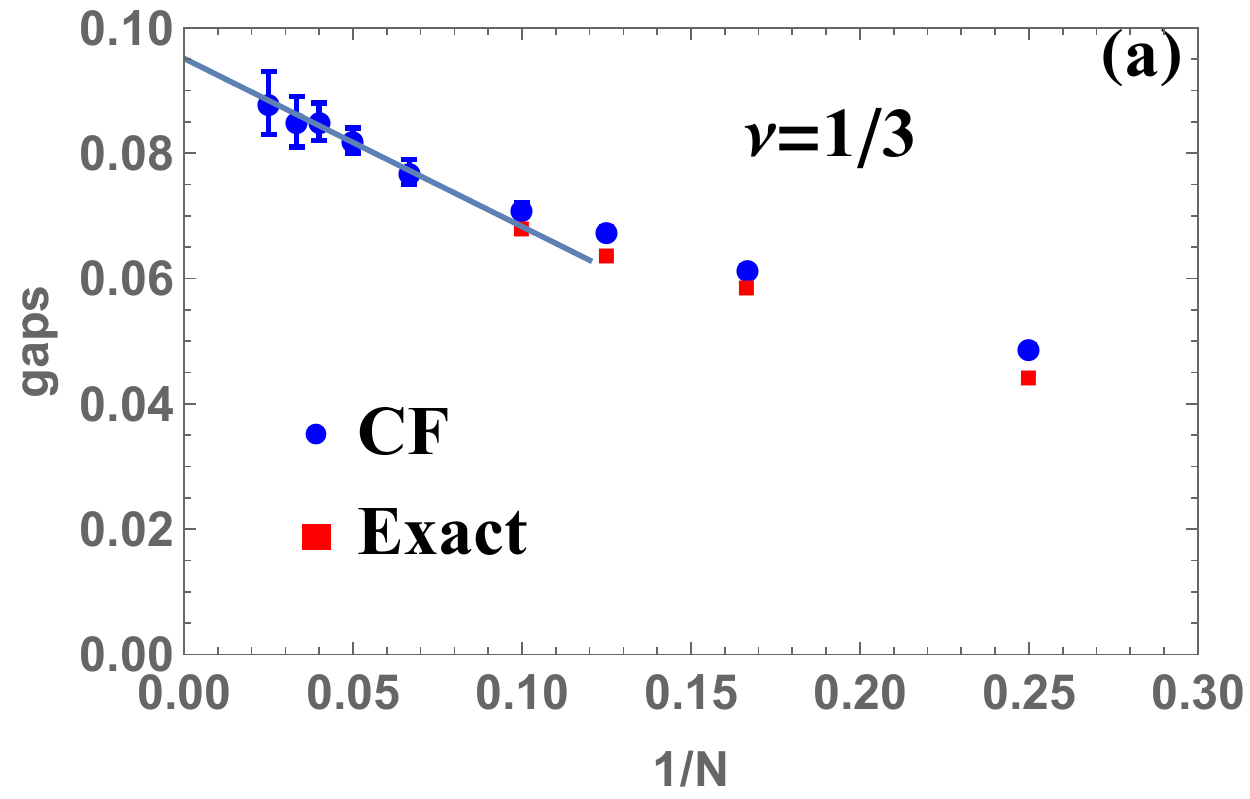} 
\includegraphics[width=7.5CM,height=5.0CM]{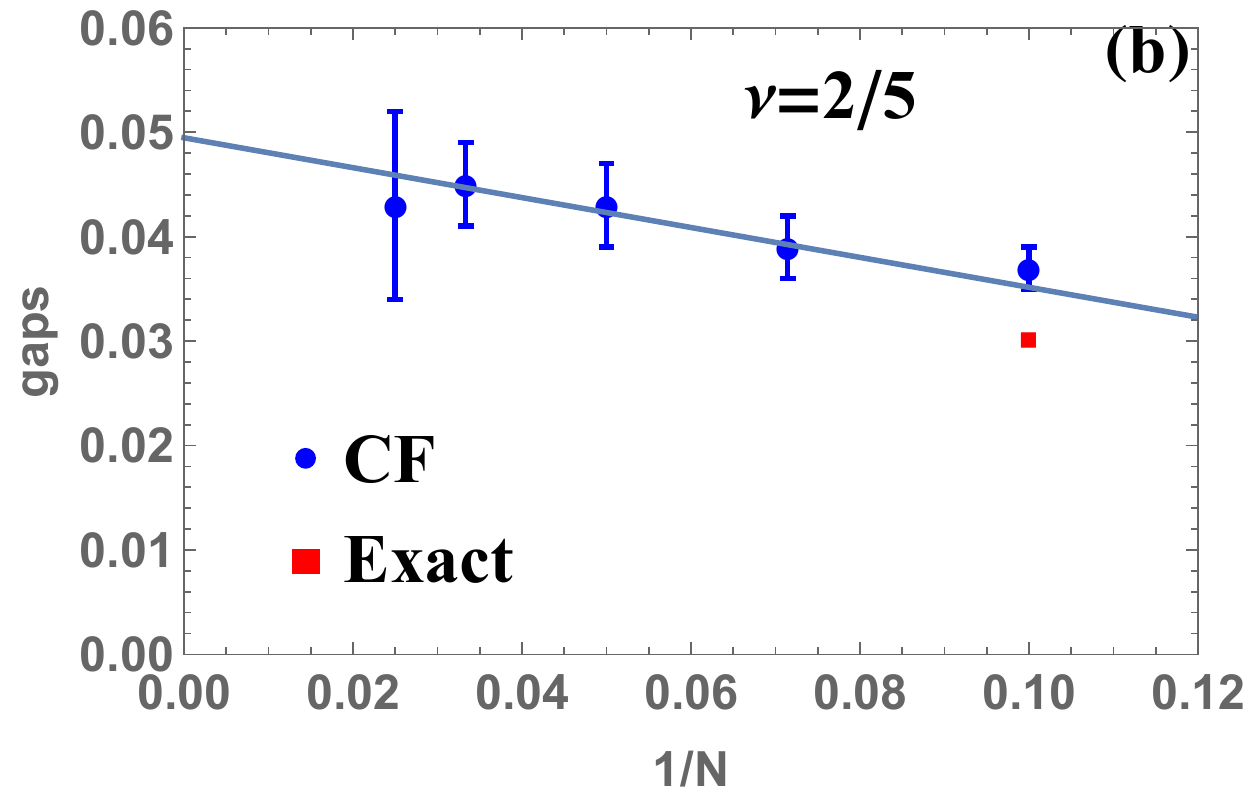} 
\end{center}
\caption{The excitation gap for (a) $\nu=1/3$ and (b) $\nu=2/5$ in units of $e^{2}/\epsilon l$.  }
\label{gapfig}
\end{figure}

We have also calculated the overlaps between our CF wave functions and exact eigenstates of Coulomb interaction. For this purpose, we calculate the inner product of the CF wave function with each Slater determinant basis function. To deal with large dimensional Hilbert spaces, we initially perform $2\times 10^5$ iterations to obtain all inner products, and then perform $1.5\times 10^6-5\times 10^6$ iterations for those basis functions whose squared inner product is larger than some number (0.001 for systems whose dimension is smaller than 1000 and 0.0001 for systems whose dimension is over 1000). The resulting overlaps for the incompressible ground states at $\nu=2/5,3/7$ and the CF quasiholes and CF quasiparticles at $\nu=2/5$ are given in Table~\ref{overlap}, along with the statistical error in the Monte Carlo evaluation of the overlap integral.

\begin{table}[!h]
\begin{tabular}{|c|c|c|c|c|}
\hline
N&$N_\phi$& state & overlap&$D$\\ \hline
2&5& $2/5$ ground state& $1.000000\pm 0.000000$&2\\ \hline
4&10& $2/5$ ground state& $0.99781\pm 0.00008$&22\\ \hline
6&15& $2/5$ ground state& $0.9967\pm 0.0002f$&335\\ \hline
8&20& $2/5$ ground state& $0.962\pm 0.002$&6310\\ \hline
3&7& $3/7$ ground state& $0.99532\pm 0.00006$&5\\ \hline
6&14&$3/7$ ground state& $0.9942\pm 0.0001$&217\\ \hline
4&11&CF quasiparticle at $1/3$&$0.9825\pm 0.0003$&30\\ \hline
5&14&CF quasiparticle at $1/3$&$0.9923\pm 0.0004$&143\\ \hline
6&17&CF quasiparticle at $1/3$&$0.979\pm 0.001$&728\\ \hline
3&8& CF quasihole at $2/5$ & $0.99887\pm 0.00003$&7\\ \hline
5&13& CF quasihole at $2/5$  & $0.9912\pm 0.0004$&99\\ \hline
7&18& CF quasihole at $2/5$  & $0.987\pm 0.002$&1768\\ \hline
5&12& CF quasiparticle at $2/5$  & $0.9979\pm 0.0001$&66\\ \hline
7&17& CF quasiparticle at $2/5$  & $0.9852\pm 0.0005$&1144\\ \hline
\end{tabular}
\caption{\label{overlap}Overlaps of CF wave functions with exact Coulomb eigenstates for several systems. The number $D$ is the Hilbert space dimension, i.e., the number of linearly independent basis states in the subspace with the relevant CM momentum quantum number.}
\end{table}

\section{Conclusions and future outlook}
\label{Conclusions}

We have succeeded in constructing LLL wave functions for composite fermions on a torus for a large class of states called proper states. These include the ground states and charged and neutral excitations at filling factors $\nu=n/(2pn+1)$, as well as all quasidegenerate ground states at arbitrary filling factors of the form $\nu=\nu^*/(2p\nu^*+1)$. These wave functions satisfy the correct boundary conditions, and are demonstrated, by explicit calculation, to be almost exact representations of the actual Coulomb ground states. The construction of these wave functions is complicated by the fact that the standard JK projection does not produce valid wave functions. The principal achievement of our work is to come up with a modified projection method that does. The resulting wave functions allow calculations for a large number of composite fermions on a torus.

Our modified LLL projection method identifies an operator $\hat{g}(\partial /\partial z,z)$ corresponding to each single particle state $f(\bar{z},z)$ such that 
\be
\Psi = e^{\sum_i\frac{z_i^2-|{z}_i|^2}{4l^2}} F_1^{2p}(Z)\chi[\hat{g}_{i}(\partial/\partial z_j,z_j) J^p_j] \nonumber
\ee
satisfies the correct boundary conditions for all proper states $\chi(f_i(\bar{z}_j,z_j))$. The rule for constructing  $\hat{g}(\partial /\partial z,z)$ is to bring all $\bar{z}$ to the left in  $f(\bar{z},z)$ and then make the replacement $\bar{z}\rightarrow 2l^2\hat{D}$, where $\hat{D}=2\partial/\partial z$ when it acts on $J^p$ and $\hat{D}=\partial/\partial z$ otherwise.  

It would be appropriate to mention certain shortcomings of our construction. As noted earlier, the LLL projection of the Jain states at $\nu=n/(2n-1)$ is difficult to evaluate. However, we expect the LLL projections of these wave functions also to be accurate, in view of the fact that the CF theory produces very accurate wave functions for these states in the disk and the spherical geometries \cite{Wu93,Suorsa11,Davenport12,Balram15b}.  As another point, we note that the proper states do not span the full LLL Hilbert space, as can be seen by simple counting for small systems.  This should be contrasted with  the construction in the disk or the spherical geometries where, by considering arbitrarily high energy excitations, the wave functions for composite fermions eventually span the entire LLL Hilbert space. This limitation is not disastrous, however, because the proper states do capture all low-energy states, including states of immediate interest, such as the incompressible FQH states and their charged and neutral excitations.

In addition to the topics mentioned in the introduction, our approach suggests a number of possible directions. One of the developments in the field of the FQHE has been to seek a connection between the FQHE physics and CFT, and, in particular, to express FQH wave functions as correlators of CFTs, with particles represented as primary fields \cite{Fubini91,Cristofano91,Cristofano91a,Cristofano91b,Moore91,Hansson07,Hansson07a}.  As mentioned in the introduction, the CFT approach has served as a guide for the construction of wave functions for composite fermions on a torus. It would be interesting to ask whether the wave functions constructed in the present work have a natural CFT representation. 

Our approach can also be generalized to construct, in the torus geometry, the unprojected parton wave functions \cite{Jain89b} and also the wave functions for composite-fermionized bosons in the lowest LL \cite{Cooper99,Viefers00,Cooper01,Regnault03,Chang05b,Regnault06,Cooper08,Viefers08}. 

Finally, we note that even though our wave functions are already very accurate, it should be possible to improve them further by allowing $\Lambda$L mixing, following similar studies in the disk and spherical geometries \cite{Jain07} that employ the method of CF diagonalization \cite{Mandal02}. It would also be interesting to investigate, as in the spherical geometry, whether certain excited states at the effective flux $N^*_\phi$ are annihilated by LLL projection during the process of composite fermionization \cite{Dev92,He94,Wu95}, and perform a counting of the remaining excited states \cite{Balram13}.

In conclusion, we expect that the ability to construct explicit wave functions for a large class of FQH states and their excitations on a torus will provide important new insight into several interesting questions for which the torus geometry is well suited. 

\section{Acknowledgment} 

This work was supported in part by the U. S. National Science Foundation, Grant No. DMR-1401636 (S.P and J.K.J), and the DFG within the Cluster of Excellence NIM (Y.H.W.).  S.P. thanks Ajit Balram for numerous helpful discussions and generous help with computer programming, Bin Wang for help on special functions, and Jie Wang for advice. We thank Ajit Balram, Mikael Fremling and Hans Hansson for valuable comments on the manuscript, and are grateful to the developers of the DiagHam codes that were used to perform exact diagonalization. We thank Di Xiao for his expert help with Fig.~1.

\begin{appendices}

\section{Certain properties of the lowest filled Landau level} \label{Laughlin wave functions}

It is clear that Laughlin's Jastrow wave function for $\nu=1$ given in Eq.~\ref{LLL filled3} is equal to the Slater determinant $\chi_1[z_i,\bar{z}_i]$ in Eq.~\ref{LLL filled2}, modulo a normalization factor. This follows because the Laughlin wave function is the unique wave function in the LLL in which each electron sees a single zero at every other particle. In this appendix we show that the two wave functions have the same behavior under CM translation. 

Let us first consider the Laughlin wave function at $\nu=1/(2p+1)$ in the torus geometry constructed previously by Haldane and Rezayi  \cite{Haldane85,Haldane85b,Greiter16}. It is given by
\begin{equation}
\label{laughlin wf with theta form}
\begin{gathered}
\Psi [z_i]=e^{\sum_{i=1}^{N} \frac{z_i^2-|z_i|^2}{4l^2}}\chi[z_i]\\
\end{gathered}
\end{equation}
with 
\begin{equation}
\label{f(z) laughlin theta}
\chi[z_i]=F(Z) \prod_{i<j}^N \left( \theta \left(\frac{z_i-z_j}{L_1}|\tau \right)\right)^{2p+1}
\ee
With the periodic boundary conditions of Eq.~\ref{pb}, $F(Z)$ should satisfy:
\begin{equation}
\label{pb F(Z) theta}
\begin{gathered}
\frac{F(Z+L_1)}{F(Z)}=(-1)^{N_\phi -(2p+1)} e^{i \phi_1}\\
\frac{F(Z+L_1\tau)}{F(Z)}=(-1)^{N_\phi -(2p+1)} e^{-i\pi (2p+1)(2Z/L_1+\tau)} e^{i \phi_\tau}
\end{gathered}
\end{equation}
The factor $F^{(n)}(Z)$ is thus an eigenfunction of the \rm{CM} translation operator. 
The solutions for a complete and orthogonal basis for $F(Z)$ are \cite{Greiter16}:
\begin{widetext}
\be
\begin{gathered}
\label{F(Z) theta}
F^{(n)}(Z)=e^{iK^{(n)}Z} \prod_{\nu =1}^{(2p+1)} \theta(Z/L_1-W_{\nu}^{(n)}|\tau)\\
K^{(n)}=(\phi_1-\pi N_\phi+2\pi n)/L_1\\
W_\nu^{(n)}=\frac{1}{2\pi (2p+1)}\left(\phi_\tau-\phi_1 \tau+\pi N_\phi \tau -\pi N_\phi-2\pi n \tau-2p\pi+(\nu-1)2\pi \right)
\end{gathered}
\ee
\end{widetext}

We now show that we can analyze the properties of $\psi_1[z_i,\bar{z}_i]$ under CM and relative translation without assuming the form of the relative part given in Eq.~\ref{f(z) laughlin theta}, but by directly using the Slater determinant form in Eq.~\ref{LLL filled2}. 

The \rm{CM} translation operator, defined as $T_{\rm{CM}}\left(\frac{L_1}{N_\phi}\right)=\prod_{i=1}^{N}T\left(\frac{L_1}{N_\phi}\right)$, translates every particle by $L_1/N_\phi$, which is the smallest translation that preserves the boundary condition:
\begin{equation}
\label{theta laughlin dege}
T_{\rm{CM}}(\frac{L_1}{N_\phi}) F^{(n)}(Z)=(-1)^{N+1}e^{i\frac{\phi_1+2\pi n}{m}} F^{(n)}(Z)
\end{equation}

First we can define the relative magnetic translation operator \cite{Haldane85b}
\be
t_i^{rel}(a)=t_i(\frac{N-1}{N}\vec{a})\prod_{j(j\neq i)}t_j(-\frac{\vec{a}}{N})
\ee
The relative magnetic translation operators only translate the relative part while keeping the CM part fixed. By considering the translation operators acting on each individual matrix element in $\chi_1[f_i(z_j)]$ expressed as the Slater determinant in Eq.~\ref{LLL filled2} and making use of Eq.~\ref{index n}, it is found that:
\be
\label{relative}
\begin{gathered}
t_i^{rel}(L_1)\psi_1[z_i,\bar{z}_i]=(-1)^{N-1}\psi_1[z_i,\bar{z}_i]\\
t_i^{rel}(L_1\tau)\psi_1[z_i,\bar{z}_i]=(-1)^{N-1}\psi_1[z_i,\bar{z}_i]
\end{gathered}
\ee
which means that $\psi_1[z_i,\bar{z}_i]$ is the eigenstate of $t_i^{rel}(L_1)$ and $t_i^{rel}(L_1\tau)$, and the eigenvalue is independent of $\phi_1$ and $\phi_\tau$.

The CM magnetic translation operators are defined as:
\be
t_{\rm{CM}}(\vec{a})=\prod_{i=1}^N t_i(\vec{a})
\ee
 By applying $t_{\rm{CM}}\left(\frac{L_1}{N_\phi}\right)$ and $t_{\rm{CM}}(L_1\tau /N_\phi)$ on $\psi_1[z_i,\bar{z_i}]$ and making use of Eq.~\ref{index n}, we find 
\be
\label{CM}
\begin{gathered}
t_{\rm{CM}}\left(\frac{L_1}{N_\phi}\right)\psi_1[z_i,\bar{z}_i]=(-1)^{N-1} e^{i\phi_1}\psi_1[z_i,\bar{z}_i]\\
t_{\rm{CM}}\left(\frac{L_1\tau}{N_\phi}\right)\psi_1[z_i,\bar{z}_i] =(-1)^{N-1}e^{i\phi_\tau}\psi_1[z_i,\bar{z}_i]
\end{gathered}
\ee
Let us now assume that the CM part of $\chi_1[f_i(z_j)]$ is $F_1(Z)$. Its form can be derived from Eq.~\ref{CM}: 
\begin{widetext}
\be
\label{pb F1 2}
\begin{gathered}
\frac{F_1(Z+L_1)}{F_1(Z)}=\frac{T_{\rm{CM}}\left(\frac{L_1}{N_\phi}\right)\chi_1}{\chi_1}=(-1)^{N-1} e^{i\phi_1}\\
\frac{F_1(Z+L_1 \tau)}{F_1(Z)}=\frac{T_{\rm{CM}}\left(\frac{L_1\tau}{N_\phi}\right)\chi_1}{\chi_1}=(-1)^{N-1}e^{i(\phi_\tau-\pi\tau-\frac{2\pi Z}{L_1})}
\end{gathered}
\ee
\end{widetext}
This is exactly the same as Eq.~\ref{pb F(Z) theta} with $p=0$. Hence the CM component $F_1(Z)$ of the Slater determinant wave function in Eq.~\ref{LLL filled2} is the same as the CM component $F(Z)$ in the Laughlin wave function of Eq.~\ref{LLL filled3}.

\section{Wave functions for filling factors $\nu=\frac{n}{2pn-1}$}
\label{Negative}

In the main body of this paper, we only discuss how to construct wave functions for the filling factors $\nu=\frac{n}{2pn+1}$. In this appendix, we show that we can construct the Jain wave functions for the filling factors $\nu=\frac{n}{2pn-1}$. The explicit evaluation of the LLL projection of these wave functions is much more difficult, however.

For filling factors $\nu=\frac{n}{2pn-1}$, the effective magnetic field for composite fermions is anti-parallel to the physical magnetic field. The relation Eq.~\ref{number of flux} does not change, but $N_\phi^*$ is a negative integer given by:
\be
N_\phi^*=-\frac{L_1^2Im(\tau)|B^*|}{\phi_0}
\ee
The single particle wave functions are obtained by complex conjugation of the wave functions given in Eq.~\ref{single wf with theta form} and Eq.~\ref{f(z)}. Below we show that by simply taking the complex conjugate of $\psi_1(z)$ and plugging it into Eq.~\ref{CF product}, we obtain a valid wave function satisfying the correct periodic boundary conditions.

The complex conjugate of $\psi_1(z)$ is (note that $l^*$ is the effective magnetic length defined in Eq.~\ref{l*}, which is a real number)
\be
\psi_1^*(z)=e^{\bar{z}^2-|z|^2\over 4l^{*2}}f_1^*(z)
\ee
where $f_1^*(z)$ satisfies 
\begin{equation}
\begin{gathered}
\label{pb f* 1}
\frac{T(L_1)f^*_1(z)}{f^*_1(z)}=e^{-i\phi_1^*}\\
\frac{T(L_1\tau)f^*_1(z)}{f^*_1(z)}=e^{-i(\phi_\tau^*-\pi |N_\phi|(2\bar{z}/L_1+\bar{\tau}))}
\end{gathered}
\end{equation}
As before, we consider the product
\ba
\psi(z,\bar{z})&=&\psi^{*}_1(z,\bar{z})\prod_i\psi^{(i)}(z,\bar{z}) \nonumber \\
&=& e^{\frac{z^2-|z|^2}{4l^{2}}}\left[e^{\bar{z}^2+z^2-2|z|^2 \over 4l^{*2}}f_1^{*}(z,\bar{z}) \prod_{i} f_1^{(i)}(z,\bar{z})\right]
\ea
in which we have used:
\be
N_\phi=-|N_\phi^*|+\sum_i N_\phi^{(i)}
\ee
By making use of Eq.~\ref{pb f* 1} and the translational properties of $e^{\bar{z}^2+z^2-2|z|^2 \over 4l^{*2}}$, it can be shown that 
\be
h(z,\bar{z})\equiv e^{\bar{z}^2+z^2-2|z|^2 \over 4l^{*2}}f_1^{*}(z,\bar{z})
\ee
 satisfies:
 \be
 \frac{T(L_1)h(z,\bar{z})}{h(z,\bar{z})}=e^{-i\phi_1^*}
 \ee
 \be
 \frac{T(L_1\tau)h(z,\bar{z})}{h(z,\bar{z})}=e^{-i(\phi_\tau^*-\pi |N_\phi|(2z/L_1+\tau))}
 \ee
 Therefore, the product $\psi(z,\bar{z})$ satisfies the correct periodic boundary conditions provided we set 
 \be
 \phi_1=-\phi_1^*+\sum_i \phi_1^{(i)}
 \ee
 \be
 \phi_\tau=-\phi_\tau^*+\sum_i \phi_\tau^{(i)}
 \ee
 However, it is difficult to explicitly obtain the projected states with negative flux attachment because $\bar{z}$ appears in the Jacobi theta functions.

\section{CM degeneracy of the Laughlin state derived from CF theory}
\label{CMLaughlin}

It is known from general considerations that the ground state at $\nu=n/(2pn+1)$ has a $(2pn+1)$-fold degeneracy arising from the CM degree of freedom. The CF theory naturally produces a single wave function at these filling factors, namely the LLL projection of $\Psi_n\Psi_1^{2p}$. We show in the following Appendix how we can derive the correct degeneracy within the CF approach. In this Appendix we consider the special case of $\nu={1\over 2p+1}$, where it is possible to display the CM degeneracy explicitly and to construct all $2p+1$ wave functions.
 
According to Eq.~\ref{CF product}, the wave function for the ground state at $\nu={1\over 2p+1}$ is given by
\be
\Psi_{1/(2p+1)}=\Psi_1^{2p+1}
\label{Laughlin-CF}
\ee
In Appendix \ref{Laughlin wave functions}, we have shown that, apart from the factor $e^{\frac{\sum_i (z_i^2-|z_i|^2)}{4l^2}}$, it is possible to write the wave function $\Psi_1$ as a product of a CM term and a wave function that depends only on relative coordinates: 
\be
\Psi_1=e^{\frac{\sum_i (z_i^2-|z_i|^2)}{4l^2}}F_1(Z)\left({\chi_1\over F_1(Z)} \right)
\ee
where $\chi_1/F_1(Z)$ depends only on the relative coordinates $z_i-z_j$.

We therefore write:
\be
\label{1/3}
\Psi_{1/(2p+1)}[z_i]= e^{\frac{\sum_i (z_i^2-|z_i|^2)}{4l^2}}F_{1/(2p+1)}(Z)\left(\frac{\chi_1}{F_1(Z)}\right)^{2p+1}
\ee
where we have allowed for a general \rm{CM} part.
Since $\chi_1$ only contains single particle wave functions in the LLL, there is no need for LLL projection. To solve for the explicit form of $F_{1/(2p+1)}(Z)$, we need to use periodic boundary conditions (setting all phase factors to be zero for convenience):
\ba
\frac{t_i(L_1)\Psi_{1/(2p+1)}[z_i,\bar{z}_i]}{\Psi_{1/(2p+1)}[z_i,\bar{z}_i]}&=&\frac{T_i(L_1)\left[\frac{F_{1/(2p+1)}(Z)}{F_1^{2p+1}(Z)}\chi_1^{2p+1}\right]}{\frac{F_{1/(2p+1)}(Z)}{F_1^{2p+1}(Z)}\chi_1^{2p+1}} \nonumber \\
&=&\frac{F_{1/(2p+1)}(Z+L_1)}{F_{1/(2p+1)}(Z)}\frac{F_1^{2p+1}(Z)}{F_1^{2p+1}(Z+L_1)} \nonumber \\
&=&(-1)^{(2p+1)(N-1)}\frac{F_{1/(2p+1)}(Z+L_1)}{F_{1/(2p+1)}(Z)}\nonumber \\
\ea
In the last line we have used the periodic property of $F_1(Z)$ given in Eqs.~\ref{CM} and \ref{pb F1 2}.

Making use of the periodic boundary conditions $t_i(L_1)\Psi_{1/(2p+1)}[z_i]=\Psi_{1/(2p+1)}[z_i]$ and $t_i(L_1\tau)\Psi_{1/(2p+1)}[z_i]=\Psi_{1/(2p+1)}[z_i]$, we have
\be
\label{pb F 1/3}
\begin{gathered}
\frac{F_{1/(2p+1)}(Z+L_1)}{F_{1/(2p+1)}(Z)}=(-1)^{(2p+1)(N-1)}\\
\frac{F_{1/(2p+1)}(Z+L_1\tau)}{F_{1/(2p+1)}(Z)}=(-1)^{(2p+1)(N-1)}e^{-i(2p+1)\pi \left(\tau+2Z/L_1\right)}\\
\end{gathered}
\ee
As shown by Haldane and Rezayi\cite{Haldane85} (also see Eq.~\ref{F(Z) theta}), there are $2p+1$ solutions to Eq.~\ref{pb F 1/3}, which demonstrates a CM degeneracy of $2p+1$. 

Furthermore, using the equation
\be
\frac{\chi_1[{f}_i(z_j)]}{F_1 (Z)}=\prod_{i<j}^N\theta \left(\frac{z_i-z_j}{L_1}|\tau \right)
\ee
it follows that
\be
\Psi_{\frac{1}{2p+1}}=e^{\frac{\sum_i (z_i^2-|z_i|^2)}{4l^2}}F_{\frac{1}{2p+1}}(Z) \left(\prod_{i<j}^N\theta \left(\frac{z_i-z_j}{L_1}|\tau \right) \right) ^{2p+1}
\ee
which is precisely the form for the Laughlin wave function derived previously\cite{Haldane85,Haldane85b,Greiter16}. The ``natural" wave function from the CF theory is that given in Eq.~\ref{Laughlin-CF}, which is a specific linear combination of the $2p+1$ degenerate ground state wave functions.

\section{CM degeneracy and CM momentum for general FQH states} \label{CM deg}

It is well known\cite{Haldane85,Haldane85b,Bernevig12} that the ground state of $\nu=\frac{p}{q}$ has a CM degeneracy of $q$ ($p$ and $q$ are relatively prime). The degenerate states can be distinguished by their CM momenta, i.e. the eigenvalues of $t_{\rm{CM}}\left(\frac{L_1}{N_\phi}\right)$.  On the other hand, as noted in the main text, the wave functions of Eq.~\ref{CF product} are, in general, not eigenstates of $t_{\rm{CM}}\left(\frac{L_1}{N_\phi}\right)$. In this section we construct degenerate ground states that have well-defined CM momenta, i.e., are eigenstates of $t_{\rm{CM}}\left(\frac{L_1}{N_\phi}\right)$, by projecting the composite fermion wave functions to corresponding momentum sectors. For simplicity, we take $\phi_1=0,\phi_\tau=0$; generalization to arbitrary boundary conditions is straightforward.

For $\nu={p\over q}$, assume that $\Psi_g$ is a ground state wave function but does not have a well-defined CM momentum. A ground state with a well-defined CM momentum $k$ ($k$ is an integer between 0 and $N_\phi-1$, but it cannot be any integer in this range, as will be explained soon) can be obtained by projecting the wave function into this momentum sector. This is accomplished most elegantly by application of the projection operator $P_k$ (due to Fremling\cite{Fremling17}):
\be
P_k=\frac{1}{\sqrt{q}}\sum_{j=0}^{q-1}\left[e^{-i2\pi{k\over N_\phi}}t_{\rm{CM}}\left(\frac{L_1}{N_\phi}\right)\right]^j
\ee
Consider the application of the CM translation operator on $P_k \Psi_g$:
\begin{widetext}
\be
t_{\rm{CM}}\left(\frac{L_1}{N_\phi}\right)P_k \Psi_g=e^{i2\pi{k\over N_\phi}}{1\over \sqrt{q}}\left(e^{-i2\pi{kq\over N_\phi}}\left[t_{\rm{CM}}\left(\frac{L_1}{N_\phi}\right)\right]^q+\sum_{j=1}^{q-1}\left[e^{-i2\pi{k\over N_\phi}}t_{\rm{CM}}\left(\frac{L_1}{N_\phi}\right)\right]^j\right)\Psi_g
\ee
\end{widetext}
Provided we have 
\be
e^{-i2\pi{kq\over N_\phi}}\left[t_{\rm{CM}}\left(\frac{L_1}{N_\phi}\right)\right]^q\Psi_g=\Psi_g
\label{condition1}
\ee
$P_k \Psi_g$ will have a well-defined CM momentum:
\be
t_{\rm{CM}}\left(\frac{L_1}{N_\phi}\right)P_k \Psi_g=e^{i2\pi{k\over N_\phi}}P_k \Psi_g
\ee
Let us now obtain the values of $k$ for which Eq.~\ref{condition1} is satisfied.

For this purpose, we need to use the fact that the eigenvalue for the operator $\left[t_{\rm{CM}}\left(\frac{L_1}{N_\phi}\right)\right]^q$ is fixed to be $(-1)^{pq(N-1)}e^{iq\vec{k}_r\cdot \vec{L}_1/N_\phi}$\cite{Haldane85b,Bernevig12}. Here $\vec{k}_r$ is the relative momentum\cite{Haldane85b,Bernevig12}:
\be
t_i^{rel}(p\vec{L}_{mn})\Psi=(-1)^{pq(N-1)}e^{-i{p\over N}\vec{k}_r\cdot \vec{L}_{mn}}\Psi
\ee
\be
\vec{L}_{mn}=m\vec{L}_1+n\vec{L}_2
\ee
($\vec{L}_{mn}=m\vec{L}_1+n\vec{L}_2$ $m$ and $n$ are integers while $\vec{L}_1$ and $\vec{L}_2$ are the two edges of parallelogram) which satisfies
\be
\label{rel m}
\vec{k}_r\cdot \vec{L}_1=2\pi r
\ee 
where $r$ is an integer. (By directly applying the relative translation operator on Eq.~\ref{CF product} it can be shown that $r=0$ for ground states of $\nu={n\over 2pn+1}$.)
These equations fix the acceptable values of $k$ to be
\be
\label{value k1}
k=r+jH; j=0,1,\ldots q-1,
\ee
if $(-1)^{pq(N-1)}=1$, and 
\be
\label{value k2}
k=r+N_\phi/2+jH; j=-{q-1\over2},-{q-1\over2}+1,\ldots {q-1\over2}
\ee
if $(-1)^{pq(N-1)}=-1$. Here $H={\rm gcd}(N,N_\phi)$ and $r$ is the number defined in Eq.~\ref{rel m}. Since this produces $q$ distinct values of $k$, we have exhausted all degenerate wave functions.

If by coincidence the amplitude of $\Psi_g$ in a certain momentum sector is zero, we can still construct the ground state in that momentum sector. We first project $\Psi_g$ to some momentum sector in which its amplitude is nonzero. Then we can boost that state to another momentum sector by application of $t_{\rm{CM}}\left(\frac{L_1\tau}{N_\phi}\right)$, because
\be
\label{degenerate j}
t_{\rm{CM}}\left(\frac{L_1}{N_\phi}\right)t_{\rm{CM}}\left(\frac{L_1\tau}{N_\phi}\right)=e^{i\frac{p}{q}}t_{\rm{CM}}\left(\frac{L_1\tau}{N_\phi}\right)t_{\rm{CM}}\left(\frac{L_1}{N_\phi}\right)
\ee
Repeated applications of $t_{\rm{CM}}\left(\frac{L_1\tau}{N_\phi}\right)$ will produce states at all possible $k$'s given in Eq.~\ref{value k1} and Eq.~\ref{value k2}.

The same projection operator $P_k$ can also be applied to CF quasiparticles and CF quasiholes to obtain wave functions with well-defined CM momenta.
Since the eigenvalue for $\left[t_{\rm{CM}}\left(\frac{L_1}{N_\phi}\right)\right]^q$ is simply 1 when $N$ and $N_\phi$ are relatively prime (which means that $p=N$ and $q=N_\phi$), the possible momenta for these ground states are
\be
k=0,1,\ldots N_\phi-1.
\ee

\section{Proof that $\chi[\hat{f}_i(\partial/\partial z_j,z_j)]$ commutes with the center-of-mass wave function $F_1^{2p}(Z)$} \label{proof that operator commute with F(Z)}

In this appendix, we show that $F_1^{2p}(Z)$ commutes with $\chi[\hat{f}_i(\partial/\partial z_j,z_j)]$ so long as the latter is a ``proper state'' defined in the introduction. This is crucial, as it serves as the starting point for the implementation of the JK projection. 

First, we transform the coordinates from $\{z_1, z_2, \ldots, z_N\}$ to $\{Z, w_1, w_2, \ldots, w_{N-1}\}$ where $Z$ is defined in Eq.~\ref{CMZ} and 
\be
\label{new coordinates2}
w_i\equiv z_i-\frac{Z}{N},\;\; i=1,2 \ldots N-1
\ee
What is the rule for LLL projection in the new coordinates? Let us recall that to accomplish LLL projection in the old coordinates $\{z_1, z_2, \ldots, z_N\}$, we keep the Gaussian factor $\exp\left(-\sum_{i=1}^N\frac{ |z_i|^2}{4l^2}\right)$ at the far left, and perform the replacement $\bar{z}_i \rightarrow 2l^2 \frac{\partial}{\partial z_i}$. With the factor $\exp\left(\sum_{i=1}^N\frac{ z_i^2-|z_i|^2}{4l^2}\right)$ at the far left, the replacement is
\be
\label{chain3}
\bar{z}_i \rightarrow 2l^2 \frac{\partial}{\partial z_i} +z_i
\ee
In addition, we need the chain rule for the derivatives:
\begin{multline}
\label{chain1}
\frac{\partial}{\partial z_i}=\frac{\partial}{\partial Z}+\frac{\partial}{\partial w_i}-\frac{1}{N}\sum_{j=1}^{N-1}\frac{\partial}{\partial w_j},\\
i=1,2 \ldots N-1
\end{multline}
\be
\label{chain2}
\frac{\partial}{\partial z_N}=\frac{\partial}{\partial Z}-\frac{1}{N}\sum_{j=1}^{N-1}\frac{\partial}{\partial w_j}
\ee

With Equations.~\ref{chain3}, \ref{chain1} and \ref{chain2} we can now derive the rule for projecting $\chi[\hat{f}_i(\partial/\partial z_j,z_j)]F_1^{2p}\prod_j J_j^{p}$ in the new coordinates $\{Z, w_1, w_2, \ldots, w_{N-1}\}$. The LLL projection corresponds to the following replacements:
\ba
\label{newpro1}
\bar{Z}&=&\sum_{i=1}^N \bar{z}_i \nonumber\\
&\rightarrow&2l^2\sum_{i=1}^N \frac{\partial}{\partial z_i}+\sum_{i=1}^N z_i \nonumber\\
&=&2Nl^2 \frac{\partial}{\partial Z}+Z
\ea
\ba
\label{newpro2}
\bar{w}_i&=&\bar{z}_i-\frac{\bar{Z}}{N} \nonumber\\
&\rightarrow&2l^2 \frac{\partial}{\partial z_i} +z_i-\frac{1}{N}\left(2l^2\sum_{i=1}^N \frac{\partial}{\partial z_i}+\sum_{i=1}^N z_i\right) \nonumber\\
&=&2l^2 \frac{\partial}{\partial w_i}+w_i
\ea
Equations.~\ref{newpro1} and \ref{newpro2} imply that if the unprojected $\chi[f_i(z_j)]$ does not depend on $\bar{Z}$, then the projected $\chi[\hat{f}_i(\partial/\partial z_j,z_j)]$ will be independent of $\partial/\partial Z$, and hence commute with $F_1^{2p}(Z)$. Below we show that $\chi[f_i(z_j)]$ is indeed independent of $\bar{Z}$ for proper states.

Let us consider, for simplicity, a proper state involving the lowest two LLs for illustration; the generalization to higher LLs follows along the same lines. 
With the new coordinates, the matrix elements in $\chi[f_n^{(m)}]$ are ($l^*$ is the effective magnetic length defined in Eq.~\ref{l*}): 
\be
f_1^{(m)}(z_i)=f_1^{(m)}(Z/N+w_i),i=1,2 \ldots N-1
\ee
\be
f_1^{(m)}(z_N)=f_1^{(m)}(Z/N-\sum_{i=1}^{N-1}w_i)
\ee
\begin{widetext}
\begin{multline}
\label{unp 2nd}
f_2^{(m)}(z_i)=\frac{\bar{Z}-Z}{\sqrt{2}Nl^*}f_1^{(m)}(Z/N+w_i)+\frac{\bar{w}_i-w_i}{\sqrt{2}l^*}f_1^{(m)}(Z/N+w_i)-\sqrt{2}l^*\left(\frac{\partial}{\partial Z}+\frac{\partial}{\partial w_i}-\frac{1}{N}\sum_{j=1}^{N-1}\frac{\partial}{\partial w_j}\right){f_1^{(m)}(Z/N+w_i)},\\
i=1,2 \ldots N-1
\end{multline}
\begin{multline}
\label{unp 2nd2}
f_2^{(m)}(z_N)=\frac{\bar{Z}-Z}{\sqrt{2}Nl^*}f_1^{(m)}(Z/N-\sum_{i=1}^{N-1}w_i)-\frac{\sum_{i=1}^{N-1}\left(\bar{w}_i-w_i\right)}{\sqrt{2}l^*}f_1^{(m)}(Z/N-\sum_{i=1}^{N-1}w_i)\\
-\sqrt{2}l^*\left(\frac{\partial}{\partial Z}-\frac{1}{N}\sum_{j=1}^{N-1}\frac{\partial}{\partial w_j}\right){f_1^{(m)}(Z/N-\sum_{i=1}^{N-1}w_i)}
\end{multline}
\end{widetext}
The first terms on the right-hand side of Eqs.~\ref{unp 2nd} and \ref{unp 2nd2}, which are the only terms containing $\bar{Z}$, are eliminated from the Slater determinant because they are proportional to the corresponding rows in the LLL. For the same reason, there is no $\bar{Z}$ dependence in $\chi[f_n^{(m)}]$ describing proper states.

\section{General derivation for $\hat{g}_n^{(m)}(z_j)$}
\label{proof modified JK}

In this appendix we show that $\hat{g}_n^{(m)}(z_j)$ exists for arbitrary LL $n$, and derive its explicit form. We show that, in general, $\hat{g}_n^{(m)}(z_j)$ can be obtained from $\hat{f}_n^{(m)}(z_j)$ by making the replacement $\partial_z\rightarrow 2\partial_z$ for the derivatives acting on the Jastrow factor $J_j$, where $\partial_z=\frac{\partial}{\partial z}$. 

The unprojected wave function $f_n^{(m)}(z)$ is:
\begin{widetext}
\be
f_n^{(m)}(z)=\left(a_f^\dagger \right)^{n-1}f_1^{(m)}(z)=\sum_{k_1=0}^{n-1}\binom{n-1}{k_1}\left({\bar{z} \over 2l^{*2}}\right)^{k_1}\left[\left(-{z \over 2l^{*2}}-\partial_z\right)^{n-1-k_1}f_1^{(m)}(z)\right] \nonumber\\
\ee
The standard replacement $\bar{z}\rightarrow 2l^2\partial_z +z$ for projection produces  for $\hat{f}_n^{(m)}(z)$ the expression:
\be
\label{fn bio}
\hat{f}_n^{(m)}(z)=\sum_{k_1=0}^{n-1}\binom{n-1}{k_1}\left({2l^2\partial_z +z \over 2l^{*2}}\right)^{k_1}\left[\left(-{z \over 2l^{*2}}-\partial_z\right)^{n-1-k_1}f_1^{(m)}(z)\right]
\ee
\end{widetext}
We should bear in mind that $\left({2l^2\partial_z +z \over 2l^{*2}}\right)^{k_1}$ acts on everything on its right while $\left(-{z \over 2l^{*2}}-\partial_z\right)^{n-1-k_1}$ only acts on $f_1^{(m)}(z)$. 

We know $\chi[\hat{f}_i(z_j)J_j^p]$ does not satisfy the periodic boundary conditions. We seek a modified wave function $\chi[\hat{g}_i(z_j)J_j^p]$ in which $\hat{g}_n^{(m)}(z)$ is obtained from $\hat{f}_n^{(m)}(z)$ by replacing all $\partial_z$'s acting on Jastrow factors by $\alpha \partial_z$'s, as shown in Eq.~\ref{unk g} for $n=2$. Let us define a new operator $\hat{D}_i$:
\be
\hat{D}_i\equiv \alpha \partial_{z_i} 
\ee
if $\hat{D}_i$ acts on $J_i^p$, and
\be
\hat{D}_i\equiv \partial_{z_i}
\ee
if it acts on anything else. Therefore, $\hat{g}_n^{(m)}(z_i)$ is
\begin{widetext}
\be
\hat{g}_n^{(m)}(z_i)\equiv \sum_{k_1=0}^{n-1}\binom{n-1}{k_1}\left({2l^2\hat{D}_i +z_i \over 2l^{*2}}\right)^{k_1}\left[\left(-{z_i \over 2l^{*2}}-\partial_{z_i}\right)^{n-1-k_1}f_1^{(m)}(z_i)\right]
\ee
\end{widetext}
Below we show that $\chi[\hat{g}_i(z_j)J_j^p]$ satisfies the periodic boundary condition with $\alpha=2$ for arbitrary $\Lambda$L. For convenience we take the phases $\phi_1=0$ and $\phi_\tau=0$.

We first note how $J_j^p$ and $f_1^{(m)}(z_i)$ change
when $z_i$ is translated by $L_1\tau$:
\be
T_i(L_1\tau)J_j^p=e^{ip\pi \left(\frac{2(z_j-z_i)}{L_1}-\tau+1\right)}J_j^p,j\neq i
\label{rule1}
\ee
\be
T_i(L_1\tau)J_i^p=\prod_{j(j\neq i)}e^{-ip\pi \left(\frac{2(z_i-z_j)}{L_1}+\tau+1\right)}J_i^p
\ee
\be
T_i(L_1\tau)f_1^{(m)}(z_i)=e^{-i\pi N_\phi^*\left(\frac{2z_i}{L_1}+\tau\right)}f_1^{(m)}(z_i)
\label{rule3}
\ee
With Eq.~\ref{fn bio}, Eqs.~\ref{rule1}-\ref{rule3}, and our replacement rule, we have:
\begin{widetext}
\begin{multline}
\label{gn trans1}
T_i(L_1\tau)\hat{g}_n^{(m)}(z_j)J_j^p=\\
e^{ip\pi \left(\frac{2(z_j-z_i)}{L_1}-\tau+1\right)}\sum_{k'_1=0}^{n-1}\sum_{k'_2=0}^{k'_1}\binom{n-1}{k'_1}\binom{k'_1}{k'_2}\left\{\left({2l^2\hat{D}_j +z_j \over 2l^{*2}}\right)^{k'_1-k'_2}\left({i2p\pi \alpha N_\phi^* \over L_1N_\phi}\right)^{k'_2}\left[\left(-{z_j \over 2l^{*2}}-\partial_{z_j}\right)^{n-1-k'_1}f_1^{(m)}(z_j)\right]J_j^p\right\},j\neq i
\end{multline}
Here we have used 
\be
\left({2l^2\hat{D}_j +z_j \over 2l^{*2}}\right)^{k_1}e^{ip\pi \left(\frac{2(z_j-z_i)}{L_1}-\tau+1\right)}=e^{ip\pi \left(\frac{2(z_j-z_i)}{L_1}-\tau+1\right)} \left(\left({2l^2\hat{D}_j +z_j \over 2l^{*2}}\right)+{i2p\pi \alpha N_\phi^* \over L_1N_\phi}\right)^{k_1}
\ee
where the exponential factor is a part of $T_i(L_1\tau)J_j^p$. 
Similarly proceeding, we get
\begin{multline}
\label{gn trans2}
T_i(L_1\tau)\hat{g}_n^{(m)}(z_i)J_i^p=e^{-i\pi N_\phi^*\left(\frac{2z_i}{L_1}+\tau\right)}\prod_{j(j\neq i)}e^{-ip\pi \left(\frac{2(z_i-z_j)}{L_1}+\tau+1\right)}\sum_{k_1=0}^{n-1}\sum_{k_2=0}^{k_1}\binom{n-1}{k_1}\binom{k_1}{k_2}\\ \left\{\left({2l^2\hat{D}_i +z_i \over 2l^{*2}}\right)^{k_1-k_2}\left[{i2\pi N_\phi^* \over L_1N_\phi}\left({N_\phi \over 2}-N_\phi^*-\alpha p(N-1)-\frac{iN_\phi \rm{Re}(\tau)}{2\rm{Im}(\tau)}\right)\right]^{k_2}\right.  \\ \left. 
\left[\sum_{k_3=0}^{n-1-k_1}\binom{n-1-k_1}{k_3}\left(-{z_i \over 2l^{*2}}-\partial_{z_i}\right)^{n-1-k_1-k_3}\left[{i2\pi N_\phi^* \over L_1N_\phi}\left({N_\phi \over 2}+\frac{iN_\phi \rm{Re}(\tau)}{2\rm{Im}(\tau)}\right)\right]^{k_3}f_1^{(m)}(z_i)\right]J_i^p\right\}
\end{multline}
\end{widetext}

In general, the Slater determinant wave function does not satisfy the correct periodic boundary conditions. We show below that for the proper states, and with the choice $\alpha=2$, most of the terms in the above sum are eliminated inside the Slater determinant in precisely the same manner as shown for $n=2$ in Sec. \ref{Modified LLL projection method}. The only terms that survive are the $k'_2=0$ term in Eq.~\ref{gn trans1} and the $k_2=0,k_3=0$ term in Eq.~\ref{gn trans2}. With these terms the full wave function satisfies the correct periodic boundary conditions.

To prove this, let us consider the terms containing the factors $$\left({2l^2\hat{D}_j +z_j \over 2l^{*2}}\right)^{k'_1-k'_2}\left[\left(-{z_j \over 2l^{*2}}-\partial_{z_j}\right)^{n-1-k'_1}f_1^{(m)}(z_j)\right]J_j^p$$ in Eq.~\ref{gn trans1} and $$\left({2l^2\hat{D}_i +z_i \over 2l^{*2}}\right)^{k_1-k_2}\left[\left(-{z_i \over 2l^{*2}}-\partial_{z_i}\right)^{n-1-k_1-k_3}f_1^{(m)}(z_i)\right]J_i^p$$ in Eq.~\ref{gn trans2}. If for $k'_1-k'_2=k_1-k_2$ and 
$n-1-k'_1=n-1-k_1-k_3$ 
their coefficients are identical, then these terms are eliminated from the Slater determinant,  because they are proportional to the corresponding rows in lower $\Lambda$Ls. 
Equality of their coefficients requires:
\begin{widetext}
\be
\label{binom eq}
\binom{n-1}{k'_1}\binom{k'_1}{k'_2}\left(p\alpha \right)^{k'_2}=\sum_{k_3=0}^{k'_2}\binom{n-1}{k_1}\binom{k_1}{k_2}\binom{n-1-k_1}{k_3}\left({N_\phi \over 2}-N_\phi^*-\alpha p(N-1)-\frac{iN_\phi \rm{Re}(\tau)}{2\rm{Im}(\tau)}\right)^{k'_2-k_3}\left({N_\phi \over 2}+\frac{iN_\phi \rm{Re}(\tau)}{2\rm{Im}(\tau)}\right)^{k_3}
\ee
\end{widetext}
By making use of the identity
\be
\binom{n-1}{k_1}\binom{k_1}{k_2}\binom{n-1-k_1}{k_3}=\binom{k'_2}{k_3}\binom{n-1}{k'_1}\binom{k'_1}{k'_2}
\ee
Eq.~\ref{binom eq} becomes:
\be
{N_\phi}-N_\phi^*-\alpha p(N-1)=\alpha p
\ee
which is exactly Eq.~\ref{alpha}, giving $\alpha=2$.

This completes the proof for the statement that by making the replacement $\partial_{z_j} \rightarrow 2\partial_{z_j}$ for operators acting on $J_j$ in $\hat{f}_n^{(m)}(z_j)$, we generate a new projection operator $\hat{g}_n^{(m)}(z_j)$ such that $\chi[\hat{g}_i(z_j)J_j^p]$ satisfies the correct periodic boundary conditions.

\section{Interaction energy}
 \label{interaction}
 
We consider a rectangle for our numerical calculation, i.e., $Re (\tau)=0$. The interaction energy must be periodic, which amounts to considering an infinite periodic expansion of the rectangle. The problem can be addressed in the following way. The usual Coulomb potential in 2D is given by
\be
V(\vec{r})=\frac{1}{r}=\int {d\vec{q}\over (2\pi)^2} \frac{2\pi}{q} e^{i\vec{q}\cdot \vec{r}}
\ee
Not being periodic, this form is not appropriate for the torus geometry. Here we use the periodic interaction \cite{Yoshioka83}
\ba
\label{V}
V(\vec{r})&=&\sum_{m,n}\frac{1}{|\vec{r}+m\vec{L}_1+n\vec{L}_2|}\\
&=&\frac{2 \pi}{L_1L_2}\sum_{\vec{q}} \frac{1}{q} e^{i\vec{q}\cdot \vec{r}}
\ea
with 
\be
\label{q}
\vec{q}=\left(\frac{2\pi m}{L_1},\frac{2\pi n}{L_2}\right)
\ee
where $\vec{L}_1$ and $\vec{L}_2$ are the edges of the rectangle, and $m$ and $n$ are integers. 
$V(\vec{r})$ satisfies the correct periodic boundary condition
\be
V(\vec{r}+m\vec{L}_1+n\vec{L}_2)=V(\vec{r})
\ee

Besides the pairwise interaction, we also need to include the self-interaction energy $W$, which represents the interaction between a particle at $\vec{r}$ and its own images at $\vec{r}+m\vec{L}_1+n\vec{L}_2$. The explicit expression for the self-interaction energy is \cite{Yoshioka83,Bonsall77}
\be
\begin{gathered}
W=-\frac{e^2}{\epsilon \sqrt{L_1L_2}}[2-\sum_{mn}^{'}\varphi_{-\frac{1}{2}}(\pi(\tau m^2+\tau^{-1}n^2))]\\
\varphi_n(z)\equiv \int_1^\infty dt e^{-zt}t^n
\end{gathered}
\ee
where the prime on the summation excludes $m=n=0$.
The interaction energy per particle for a system of N particles is then given by
\be
E=W+\frac{1}{N}\frac{2 \pi}{L_1L_2}\sum_{i<j}\sum_{\vec{q}\neq 0} \frac{1}{q} e^{i\vec{q}\cdot (\vec{r_i}-\vec{r_j})}
\ee
The $\vec{q}=0$ term is omitted as it is exactly canceled by the background-background and electron-background interactions. 

The infinite sum $\sum_{\vec{q}\neq 0}$ is convergent, as can be proven by writing out the second quantization form of Eq.~\ref{V}, and finding that each term in the sum over $\vec{q}$ is proportional to $e^{-q^2l^2}$. In our Monte Carlo programs, we truncate the sum in Eq.~\ref{q}, keeping only the terms with $|m|\leq \rm{cutoff}$, $|n| \leq \rm{cutoff}$. In Fig.~\ref{cutoff-fig} and Table~\ref{cutoff} we show the cutoff dependence of the energy for various systems. We find that the energies have converged, within our Monte Carlo uncertainty, so long as the cutoff is greater than $10$. In practice, we take the value of the cutoff to be $20$.

\begin{figure}[!h]
\begin{center}
\includegraphics[width=7.5CM,height=5.0CM]{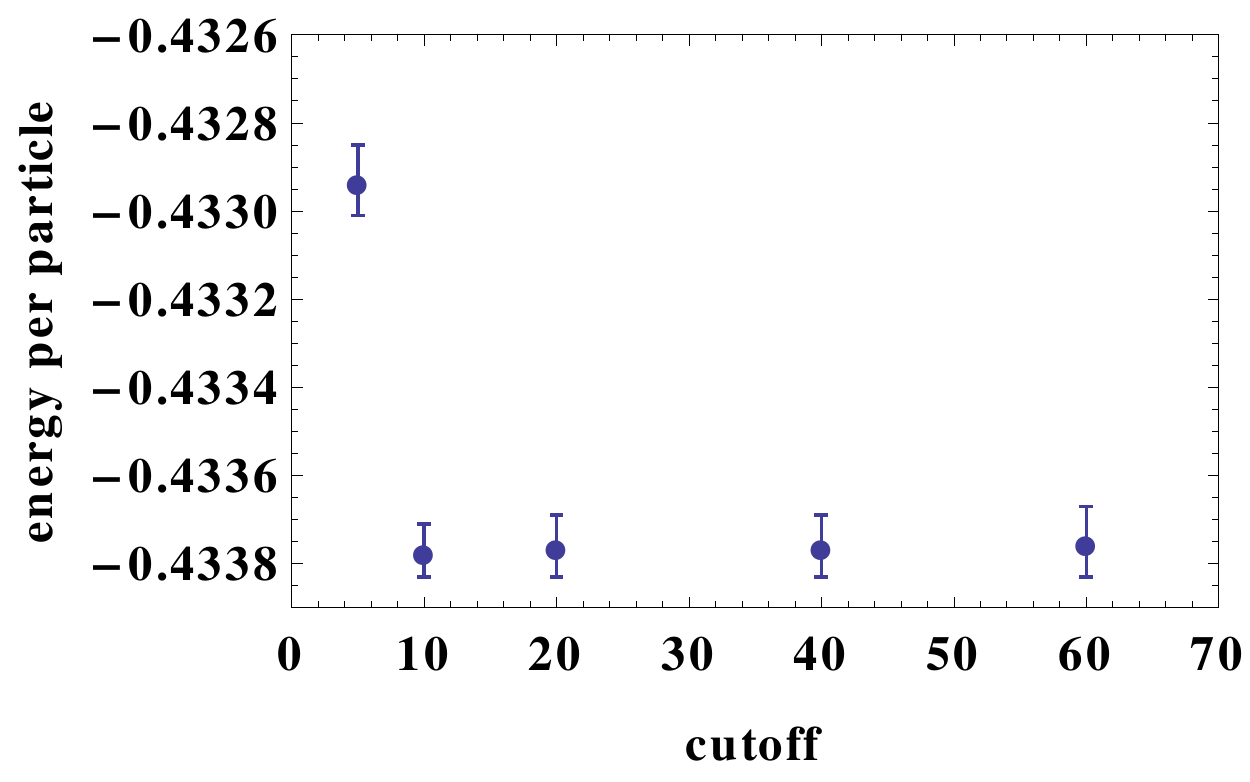} 
\end{center}
\caption{The ground state energy for $N=10,N_\phi=25$ as a function of cutoff.  The cutoff is the largest value of $|m|$ and $|n|$, where $m$ and $n$ are defined through $\vec{q}=\left(\frac{2\pi m}{L_1},\frac{2\pi n}{L_2}\right)$.}
\label{cutoff-fig}
\end{figure}

\begin{table}[!h]
\begin{tabular}{|c|c|c|c|}
\hline
N&$N_\phi$&$\rm{cutoff}=20$&$\rm{cutoff}=40$\\ \hline
10&25&$-0.43376 \pm0.00007$&$-0.43376 \pm0.00007$\\ \hline
20&50&$-0.43306 \pm0.00008$&$-0.43304 \pm0.00008$\\ \hline
20&60&$-0.41005 \pm0.00005$&$-0.41007 \pm0.00006$\\ \hline
40&120&$-0.40983\pm0.00006$&$-0.4098 \pm0.0001$\\ \hline
\end{tabular}
\caption{\label{cutoff} The ground state energies for several systems with two different values of the cutoff.}
\end{table}

We mention certain technical details that may be useful for someone who wishes to implement our method. For the evaluation of $\theta(z|\tau)$, we use the code from mymathlib modified to expand the range of $z$ to the entire complex plane. There are certain analytical formulas for the derivatives of the theta functions, but we have found it more efficient to evaluate them numerically, using 
$f'(x)=\frac{f(x+dx)-f(x-dx)}{2dx}$
and 
$f''(x)=\frac{f(x+dx)+f(x-dx)-2f(x)}{dx^2}$, and determining the optimal value of $dx$ 
by checking that Eq.~\ref{projected wf} satisfies the periodic boundary conditions as accurately as possible. We have found that the optimal value is $dx\sim 10^{-6}l-10^{-7}l$ when only the first derivatives are involved, and $dx\sim 10^{-4}l$ when second derivatives are also involved. (Here $l$ is the physical magnetic length.) We have run $\sim 5 \times 10^{6}$ Monte Carlo iterations for most of our results.

\end{appendices}

\end{document}